\documentclass[usenatbib,usegraphicx]{mn2e}
\onecolumn
\usepackage{subfigure}
\usepackage{longtable}
\usepackage{graphicx}
\usepackage{graphics}
\usepackage{keyval}
\usepackage{trig}

\def\beq{\begin{equation}}
\def\eeq{\end{equation}}
\def\bey{\begin{eqnarray}}
\def\eey{\end{eqnarray}}

\def\msun{M_\odot}
\def\sun{\odot}

\def\kms{\, {\rm km \, s}^{-1} }

\def\a0{$a_0$}

\title[Analysis of galactic tides and stars on CDM microhalos]{Analysis of galactic tides and stars on CDM microhalos}
\author[G. W. Angus and H. S. Zhao]{G. W. Angus$^{1}$\thanks{email:
gwa2@st-andrews.ac.uk} and H. S. Zhao$^{1}$ \\
$^{1}$SUPA, School of Physics and Astronomy, Univ. of St Andrews, Scotland KY16 9SS\\ National Astronomical Observatories, Chinese Academy of Sciences, Beijing 10012, PRC}

\begin{document}

\date{Accepted . Received;  in original form}
\pagerange{\pageref{firstpage}--\pageref{lastpage}} \pubyear{2006}

\maketitle
\label{firstpage}

\begin{abstract}
A special purpose N-body simulation has been built to understand the
tidal heating of the smallest dark matter substructures
($10^{-6}\msun$ and 0.01pc) from the grainy potential of the Milky Way
due to individual stars in the disk and the bulge.  To test the method
we first run simulations of single encounters of microhalos with an
isolated star, and compare with analytical predictions of the dark
particle bound fraction as a function of impact parameter. We then follow the
orbits of a set of microhalos in a realistic flattened Milky Way
potential.  We concentrate on (detectable) microhalos passing near the
Sun with a range of pericenter and apocenter. Stellar perturbers  near the orbital path of a microhalo would exert stochachstic impulses, which we apply in a Monte Carlo fashion according to the Besancon
model for the distribution of stars of different masses and ages in our Galaxy.
Also incorporated are the usual pericenter tidal heating and disk-shocking
heating.  We give a detailed diagnosis of typical microhalos and find
microhalos with internal tangential anisotropy are slightly more
robust than the ones with radial anisotropy. In addition, the dark particles generally go through of a random walk in
velocity space and diffuse out of the microhalos.
 We show that the typical
destruction time scales are strongly correlated with the stellar
density averaged along a microhalo's orbit over the age of the stellar disk. We also present the morphology of a microhalo at several epochs which may hold the key to dark matter detections.
\end{abstract}

\section{Introduction}
Cold Dark Matter (CDM) has enjoyed a great deal of success in
explaining the fluctuations of the cosmic microwave background and
large scale structure in general (Spergel et al. 2006). However, its
predictions on small scales have not been fully understood or
confirmed by observations yet.  This is partly limited by our computing
power in resolving the smallest dark matter structures and detailed
gas and stellar dynamics in structure formation.  Nevertheless in a
pioneering work Diemand, Moore \& Stadel (2005) were able to resolve a
galaxy-size dark halo down to the free-streaming scale.  The smallest
structures (called microhalos) have a minimum mass ($m_0\sim
10^{-6}M_{\sun}$), which were followed in simulations from redshift
$z=64$ to $z=26$.  As much as 0.1 percent of the dark matter in the
universe are in these substructures.  At later redshift larger
substructures collapse out of perturbations on larger scales, and
hierarchical halo formation (Lacey \& Cole 1994) predicts that the
microhalos also fall into these larger structures and are expected to
survive inside given their high internal density.  Extrapolations show
the Milky Way may contain around $10^{15}$ such microhalos with the
nearest one being on the average about 0.1 pc distant from the Earth,
and with typical size 0.01 pc.  The numerical simulations of Diemand
et al (2005, 2006) also confirm the analytical predictions of
Berezinsky, Dokuchaev \& Eroshenko (2005).

There has been much intrigue over tidal mass-loss from satellite
galaxies, for instance the tidal streams of the Sagittarius dwarf
galaxy (e.g. Savage, Freese \& Gondolo 2006) and the Magellanic clouds (e.g. Johnston, Spergel \& Haydn 2002). However, details
of the mass-loss from the smallest scale dark matter halos are still
sketchy and although some work has been done to develop
semi-analytical models of microhalo destruction (Zhao et al. 2006,
Berezinski et al. 2005, Green \& Goodwin 2006), virtually no rigorous
modelling using N-body codes has been published. It has been noted
(Green \& Goodwin 2006) that making analyses of disruption rates from
seperate consideration of the tides, stellar encounters and disk
shocking is insufficient and one consistent approach is needed. Here
we performed N-body simulations in a respectable star distribution
(Robin, Reyle \& Picaud 2003) with microhalos on orbits in a disk
type galaxy potential with dark halo, disk and bulge components,
adding contributions from star encounters in each time step over the age of the stellar disk.

The paper is laid out as follows; in \S \ref{sec:incon} we give an
analytical prescription of a typical microhalo using the NFW profile
(Navarro et al. 1997). We have two sets of initial conditions for the
internal particles of the microhalo. The first is for randomly
inclined circular orbits and the second is with a radially anisotropic
dispersion.  Then we set up the geometry of single encounters with
stars and compare the impulse approximation with numerical simulations
using the N-body tree code of Vine \& Sigurdsson (1998).

Using N-body simulations we calculated, for four typical stellar
masses (0.1, 0.3, 0.6 and 1.0 $M_{\sun}$), (i) the likely safe impact
parameter between stars and microhalos where no mass should be
stripped, (ii) the critical impact parameter below which the microhalo
will be destroyed (i.e. 95\% of particles given escape
velocity). (iii) We show for single encounters the difference between
instantaneous bound fraction of microhalo mass and the fraction
remaining bound after particle escape is allowed and the potential
shallows. Finally (iv) we present the bound fraction of microhalo mass
as a function of impact parameter for a single encounter, comparing
the microhalos with radially anisotropic and randomly oriented
circular initial conditions.

In \S \ref{sec:realsim} we present a procedure to use the Besanson
model (Robin et al. 2003) of the Milky Way to simulate the encounters
of microhalos with stars on typical orbits in the Milky Way in a three
component (bulge, disk, halo) potential over a several Gyr orbital
timescale until the microhalo is destroyed. Then in \S \ref{sec:res}
we use this model to produce some interesting results. (i) That the
survival time of microhalos and the averaged stellar density along the
orbit can be related by a power law and (ii) That the survival time is
keenly linked to the orbital energy and angular momentum. (iii)
Particles escape via diffusion during disk crossings and (iv) the
microstreams created from destroyed microhalos remain filamentary for
many Gyr. We conclude and discuss our results in \S \ref{sec:conc}.

\section{Background Theory}
\protect\label{sec:incon}
\subsection{Possible models of microhalos}
\begin{figure*}
\def\subfigtopskip{0pt}    
\def\subfigbottomskip{4pt}
\def\subfigcapskip{1pt}
\centering

\begin{tabular}{cc}
\subfigure[]{\label{fig:sigr}
\includegraphics[angle=0,width=8.5cm]{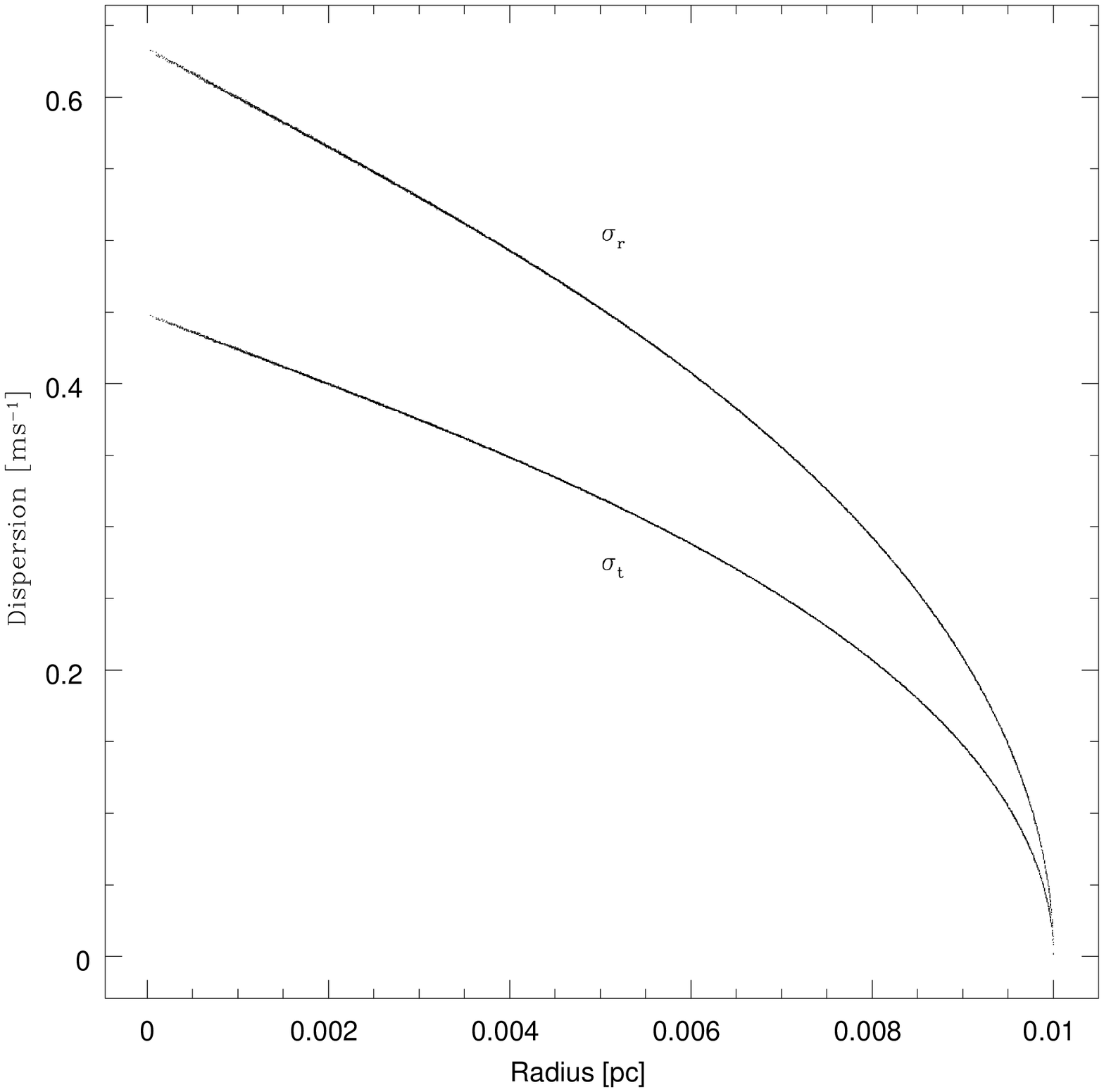}
}
&
\subfigure[]{\label{fig:vel20}
\includegraphics[angle=0,width=8.5cm]{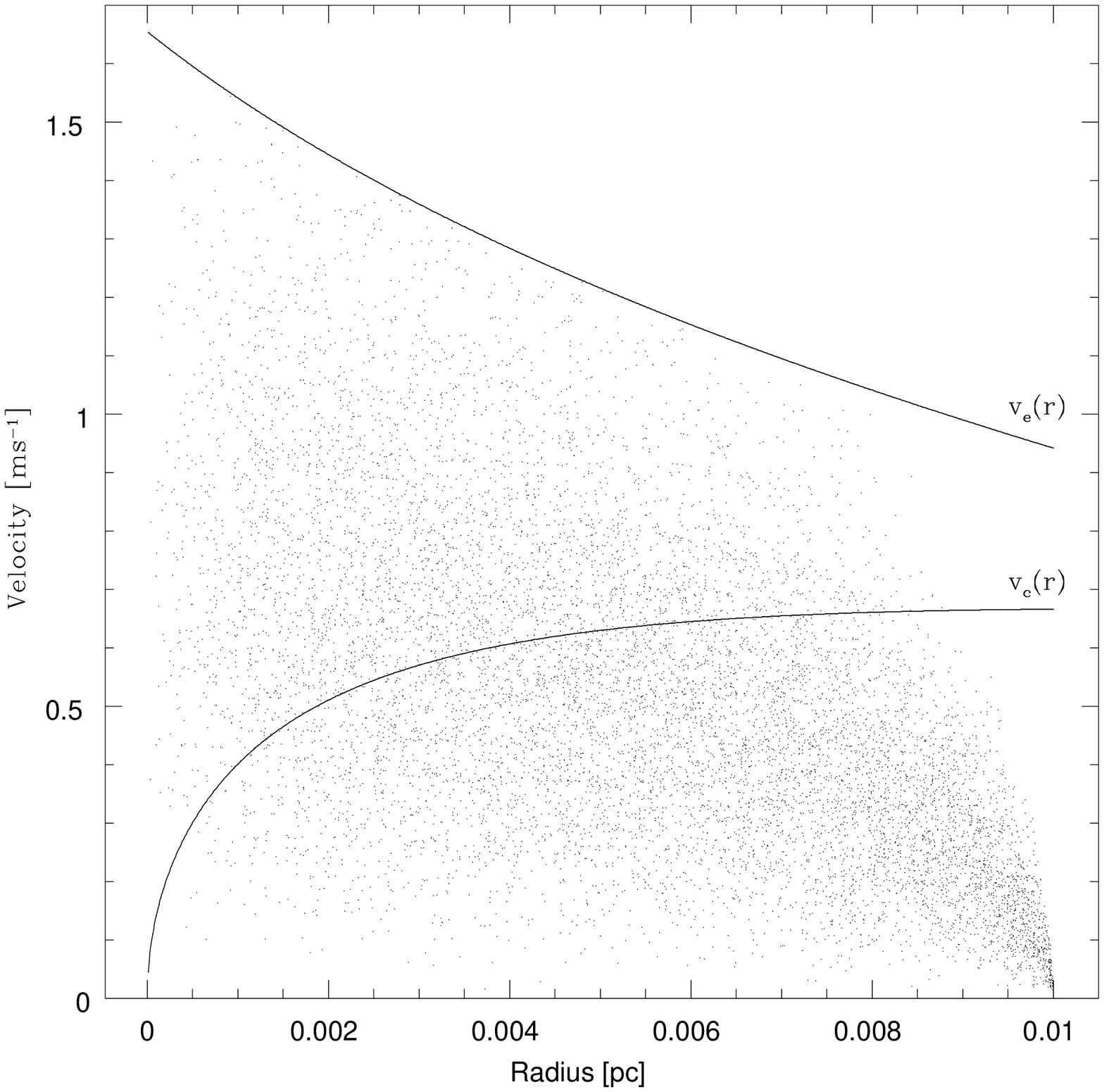}
}
\end{tabular}
\caption{(a) Shows the numerically integrated dispersions in the radial and transverse directions for the RA microhalo.(b) Shows the velocity-radius distribution for the TA and RA microhalos. The RA distribution is the scatter of particles bounded by the escape velocity (the upper, falling curve for both distributions) whereas the TA distribution follows the lower, rising curve i.e. the circular velocity.}
\end{figure*}

According to Berezinsky et al. (2005) and Diemand et al. (2005), the likely minimum mass of a microhalo set by free-streaming and collisional damping is of the order of an Earth mass, $m_o \approx 10^{-6}M_{\sun}$) and can be distributed as an NFW model with a density
\begin{equation}
\rho(r) = {A \over 4\pi} m_0 r^{-1} \left(r+r_v c^{-1}\right)^{-2}, {\rm } A^{-1}=\ln(1+c) - {c \over 1+c}
\end{equation}
where the concentration constant $c=2$ and the virial radius $r_v = 0.01pc$. The enclosed mass goes as
\begin{equation}
m(r) = A m_o \left[ \ln \left(1 + {cr \over r_v} \right) - 1 + \left(1 + {cr \over r_v}\right)^{-1}  \right] 
\end{equation}
and the potential $\phi$ or the escape velocity $V_e$ of the system goes as
\bey
\label{eqn:esc}
\nonumber
-\phi &=& {V_e^2 \over 2}\\
&=&\left[{G A m_o \over r} \ln \left(1+ {cr \over r_v}\right) - {G A m_o \over r_v} \ln (1 + c)\right]_{r<r_v}+{G m_0 \over (r,r_v)_{\rm max}}.
\eey
So the escape velocities at the edge, $r_v$ and centre respectively are
\begin{equation}
V_e (r_v) = \sqrt{2 G m_o \over r_v} \approx 0.94{\rm m\, s}^{-1}, \qquad V_e(0) \approx 1.64{\rm m\, s}^{-1}
\end{equation}
For one model we assume all dark particles are on randomly oriented tangential orbits (hereafter TA microhalos)
\begin{equation}
V_c^2 = {G m(r) \over r} $ which gives, $ V_c(r_v) \approx 0.67{\rm m\, s}^{-1}
\end{equation}
For our second model we also distribute the particles with an NFW profile, but we give the particles dispersion with an anisotropy parameter,
$\beta=1-{\sigma_{\theta}^2 \over \sigma_r^2}$=1/2  (radially anisotropic microhalos, hereafter RA microhalos). Numerical integration of the Jeans equation (see e.g. Lokas \& Mamon 2001) via the equation
\beq
2\sigma_{\theta}^2 = 2\sigma_{\phi}^2=\sigma_r^2(r)
={1 \over r \rho} \int_r^{r_v} V_{\rm c}^2 \rho dr.
\eeq
 yields dispersions (Fig. \ref{fig:sigr}) and a velocity-radius plot (Fig. \ref{fig:vel20}).  It was noted by Kazantzidis et al. (2004) that anisotropy can affect mass loss, stemming from the fact that anisotropic distributions have more particles on radial orbits that extend beyond the virial radius even though they are bound. Our two distributions (TA and RA microhalos) are unlikely to be typical for microhalos, however, they should {\em bracket} the two extreme regimes of anisotropy, thus gleaning us a lot of insight.

It is useful to estimate a few time scales, helpful for understanding the
physics of the tide.
The dynamical time of the microhalo is of order $r_v/V_c = 15$Myr.   A
typical orbit of a microhalo
around the Galaxy has an orbital timescale $\sim$ $3kpc/200 \kms$ $\sim$
150Myr.  A typical crossing of
the thin disk of the Galaxy takes a timescale of $300pc/100 \kms$ $\sim$
3Myr (depending on inclination). A penetrative
encounter with a passing star typically takes $0.02pc/200 \kms$ $\sim$
0.01Myr.  Clearly an encounter with a
star is in the impulse regime to an excellent approximation and the
pericentric tide by the Galactic potential
is almost adiabatic, and can steadily strip mass.  Disk shocking plays some
role. In the following we will concentrate
on encounters with stars.

\subsection{Impulse Approximation}
\protect\label{sec:impapp}
The impulse approximation of a massive point object's (a star's) effect on an extended body's (the microhalo's) internal particles, predicts the "kick" in velocity every internal particle
experiences due to the gravitational force the star imposes on it as they move passed each other. The analysis is based on Binney \& Tremaine (1987; p437-8) with a cartesian set of
coordinates.

\begin{figure}
\includegraphics[angle=0,width=85mm]{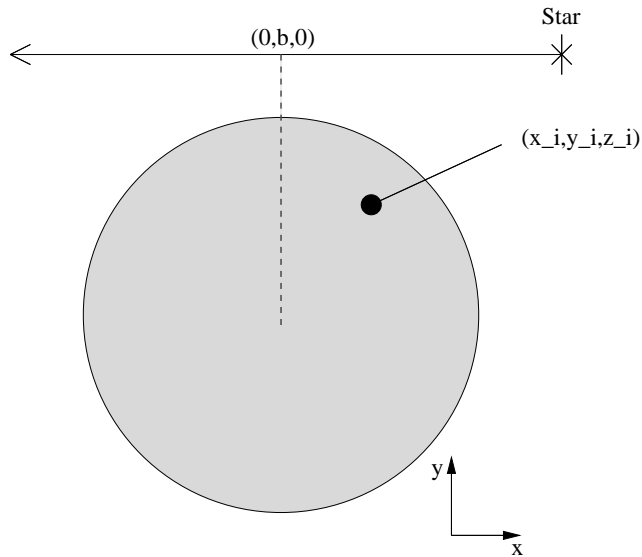}
\caption{Shows the geometry of an encounter between a star and a microhalo. Highlighted are the relative positions from Eq.\ref{eqn:imp} to the microhalo.}
\protect\label{fig:diag1}
\end{figure}

Suppose, like in Fig.\ref{fig:diag1}, we have a stationary microhalo with its centre of
mass (COM) at the origin and a star is moving from (+$\infty$,0,0) with velocity along the negative x-axis of (-$V_{\star}$,0,0). We arbitrarily define the closest approach to the microhalo's COM as (0,b,0). Here b is called the {\it impact parameter}.

The orbital speeds of particles inside the microhalo are of order 1$ms^{-1}$, therefore, during the time period which the star, with relative velocity
of $V_{rel} \sim 200\kms$, crosses the microhalo of diameter 0.02pc, the constituent particles will have made negligible movement on their respective orbits. So we can
think of them as {\it stationary} particles. The impulse acts on the y and z components of the particles' velocities, but from symmetry we see no net impulse in the x component.
Additionally, the COM velocity receives a kick in the positive y direction due to the asymmetry.

The equation that governs this impulse for the $i^{th}$ particle in the halo is
\begin{equation}
\label{eqn:imp}
\Delta V(x_i,y_i,z_i) = \frac{2GM_{\star}}{ \left[(b-y_i)^2 +{z_i}^2 \right]|V_{rel}|}(0,b-y_i,-z_i)
\end{equation}
Where $V_{rel}$ is the relative velocity between the star and microhalo and $M_{\star}$ is the mass of the perturbing star. b is the distance from the star to the microhalo's COM at
closest approach and $x_i$, $y_i$ and $z_i$ are the positions of the $i^{th}$ particle relative to the COM. 

\subsection{Comparison of Simulations with Impulse Approximation}
\protect\label{sec:coswia}

In the simplest case, in order to destroy a pristine microhalo we can set it on a collision course with a star of mass $M_*$. The gravitational attraction between the star and the
individual particles of the microhalo boosts the velocities of the particles and if the velocities w.r.t. the mean velocity are increased beyond the escape velocities (c.f. Eq.\ref{eqn:esc}) then the microhalo will
be destroyed. We used numerical simulations to gain an accurate representation of the bound fraction of mass of a microhalo after a single impact as a function of impact parameter for 4 masses of perturbers; 0.1, 0.3, 0.6 and 1.0$M_{\sun}$ stars.

In all our simulations we use time-steps of $\Delta t$=0.15Myr, $10^5$ particles, gravitiational softening of $\epsilon=10^{-4}pc$ and tree opening angle of $\theta$=0.6.

To find a rigorous correlation between bound fraction and impact parameter we configured an N-body simulation with the geometry that can be seen in Fig.\ref{fig:diag1}. To save processing time with minimal loss of accuracy, we had the star moving from 0.1 $\to$ -0.1 pc in 133 time steps (100yr) instead of from +$\infty$ $\to$ -$\infty$. We used $10^5$ particles to simulate the microhalo with initial conditions exactly as in \S \ref{sec:incon}.

After an encounter, the velocities of the individual particles have all changed and as a result, the microhalo has received a net velocity shift. We calculate every particles' velocity w.r.t. the mean velocity and then calculate the local potential for each particle imposed by other particles. Using Eq.\ref{eqn:esc} we can decide whether a particles velocity is larger than escape velocity, hence whether it is bound or not.

The issues with this method are that it makes no predictions for the future bound fraction if the microhalo is left in isolation. After a few dynamical times, the unbound particles leave the system which shallows the potential allowing other particles to escape. Fortunately, if we use the impulse approximation to simulate a stellar encounters, we can run the simulation with this as the initial conditions and safely run longer time steps which allows the particles to physically escape and we wait until the bound fraction saturates i.e. when the system reaches equilibrium.
\begin{figure}
\includegraphics[angle=0,width=85mm]{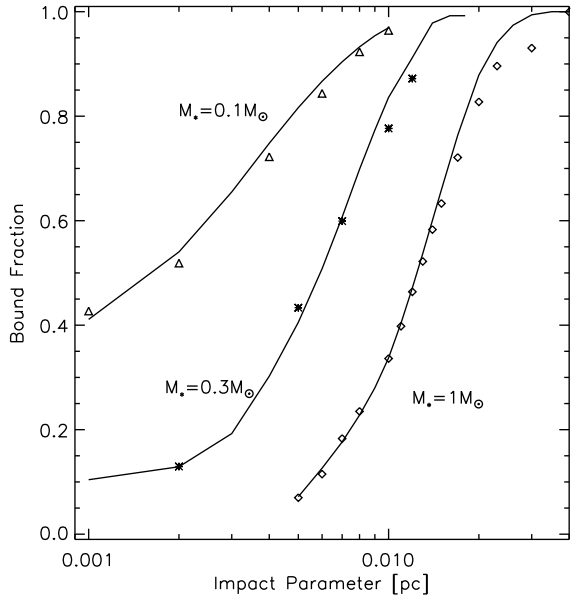}
\caption{Using TA microhalos compares bound fraction as a function of impact parameter for impulse approximation (symbols) and simulation of encounter (line) for 0.1 (triangles), 0.3 (stars) and 1.0$M_{\sun}$ (diamonds).}
\protect\label{fig:fit}
\end{figure}

We compared the simulations with the use of the impulse approximation for the TA microhalo case to see if they both gave the same results for microhalo disruptions due to stellar encounters. We simply plotted each's estimate of the instantaneous bound fraction of particles after an encounter for three different masses and a range of impact parameters (Fig.\ref{fig:fit}). Although what is plotted is the instantaneous bound fraction immediately after the impulse, this is still a good measure of how well the two methods are in agreement since comparing individual velocities is an over complicated
business and this way requires no scaling. In any case, by considering escape fraction, we are in effect comparing velocities (Eq.\ref{eqn:esc}).

Fig.\ref{fig:fit} shows there is good agreement between both estimates. This tells us two things i) that the impulse approximation is an excellent one for this system and ii) we can analytically input encounters between the microhalo and stars into our N-body simulations using the
impulse approximation. This gives us far more freedom when trying to develop a realistic evolution of microhalo structure as they orbit the Galaxy.

\subsection{Relations between impact parameter, mass of perturber and bound fraction}
For the purposes of understanding the importance of the dense star fields of the bulge and disk for destroying microhalos, it is instructive to see what impact parameters given a certain mass of a star are devastating and vice-versa.

\begin{figure}
\includegraphics[angle=0,width=85mm]{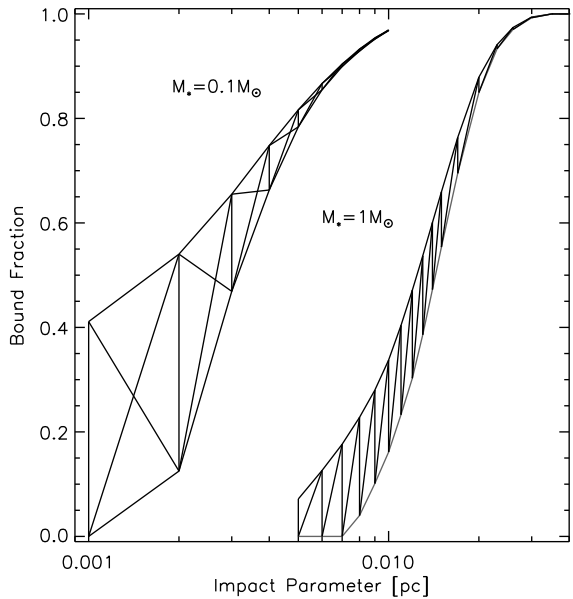}
\caption{Shows the discrepancy between the bound fraction directly after a single impulse (upper curves of the shaded regions) and after the system has been allowed to relax (lower curves) for 0.1 (shaded region to the left) and 1.0 $M_{\sun}$ (to the right) perturbing stars and a series of impact parameters. Uses the TA microhalos.}
\protect\label{fig:disc}
\end{figure}

Plotted in Fig.\ref{fig:disc} for the TA microhalos is the final bound fraction against impact parameter for 0.1 and 1.0 $M_{\sun}$ stars. Also plotted on this figure is the instantaneous bound fraction immediately after an encounter when no particles have had time to escape. The deviations are quite large, but only over a small range of impact parameter.

\begin{figure}
\includegraphics[angle=0,width=85mm]{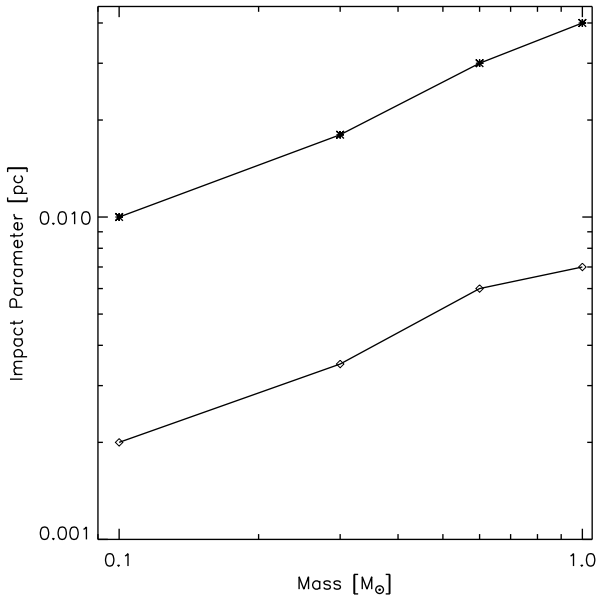}
\caption{Compares the safe impact parameter (stars) and critical impact parameter (diamonds) for four perturbing masses. Indistinguishable for TA or RA microhalos.}
\protect\label{fig:safecrit}
\end{figure}

We calculate the final bound fraction by analytically inserting a kick from a star with the impulse approximation. We run time-steps of 0.15Myr and stop the simulation when the bound fraction saturates.
In Fig.\ref{fig:safecrit} we plot the critical impact parameters for our four star masses and also plotted is the impact parameter where no mass will be lost. The trend for both appears linear over the given mass range. Critical impact parameter refers to the impact parameter for which the microhalo will lose more than 95\% of its particles.
This clearly shows how the severity of an encounter varies
with impact parameter and mass and that the distinction between a destructive encounter and a passive one is the difference of a few 100$^{ths}$ of a parsec.
\begin{figure}
\includegraphics[angle=0,width=85mm]{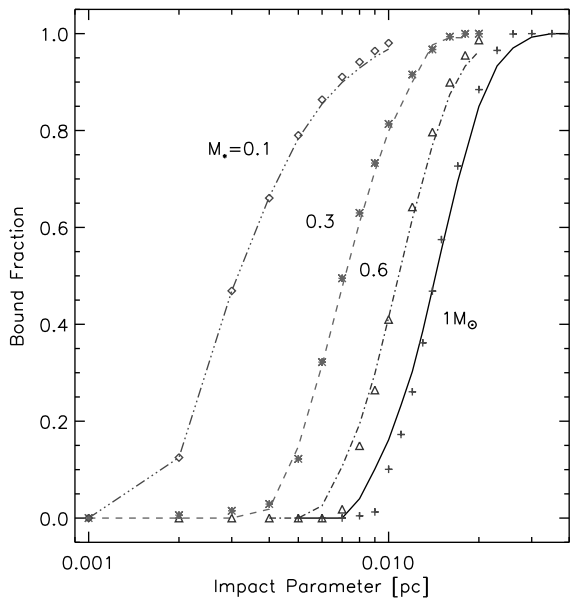}
\caption{Compares the relaxed bound fraction after single encounters with four masses of stars for TA (lines) and RA (symbols) microhalos. From left to right, the star masses are 0.1, 0.3, 0.6 and 1.0 $M_{\sun}$. Note only the slight differences in the robustness of the two types of microhalos.}
\protect\label{fig:radiso}
\end{figure}
To compare the bound fractions for our two sets of initial conditions, we have plotted final bound fraction against impact parameter for both (Fig. \ref{fig:radiso}). Although the differences are small, there is a clear trend for the RA microhalos to be slightly more susceptable to weak encounters and conversely, for TA microhalos to be more susceptable to strong encounters.
Of course, these results are only true for pristine microhalos, however, they do introduce the likely distances involved in destroying microhalos which should be transferrable to
evolved microhalos. All results were checked with fewer particles to test for convergence.
\section{Realistic Simulations}
\protect\label{sec:realsim}
As was mentioned in the introduction, there has been no consistent analysis of the combined effects of stellar encounters, disk shocking and galactic tidal stripping in one single calculation. Green \& Goodwin (2006) have come closest by using single encounter results compounded with some crude stellar mass selections for either halo or disk orbits. Here we cover all these bases consistently.

The dynamical time of the microhalo is $\sim$ 15Myr whereas the crossing time of the star across the microhalo is $\sim$ $0.02 \over 200$$pc \over {\rm km\, s}^{-1}$ = 100yr.
Therefore, it would require many billions of time steps to analyse the evolution of the microhalo in this system. However, since we know the impulse approximation is a very good one (Fig.\ref{fig:fit}), we
can run time steps of 1\% of the dynamical time and add impulses in the simulation using the impulse approximation without losing virtually any accuracy. We made specific modifications to the
N-body tree code developed by Vine \& Sigurdsson (1998) to integrate the orbits of the microhalos around a Galactic potential with disk, bulge and halo components (Helmi 2004). The logarithmic dark halo has potential
\begin{equation}
\Phi_{halo} = {1 \over 2}V_o^2 \ln \left (R^2+z^2/q^2+d^2\right).
\protect\label{eqn:galpot1}
\end{equation}
For the Miyamoto-Nagai disk
\begin{equation}
\Phi_{disk} = -{GM_{disk} \over \sqrt{R^2+(a_d+\sqrt{z^2+b_d^2})^2}}
\protect\label{eqn:galpot2}
\end{equation}
and for the spherical Hernquist bulge
\begin{equation}
\Phi_{bulge} = -{GM_{bulge} \over r+c_b}
\protect\label{eqn:galpot3}
\end{equation}
Where R, r and z are the standard representations of polar coordinates. $q^2=0.8$ shows the deviation from spherical geometry (flattening). The parameters used to fit the potential of the Milky Way to the rotation curves of stars are unchanged from Helmi (2004) i.e. $V_o$ = 186$\kms$.

\begin{table*}
\caption{Mass formulae (cf. Eq. 15) and Star Numbers for different components of the Galaxy}
\protect\label{tab:reg}
\vspace{3mm}
\begin{tabular}{lccccc}
\hline
Regime&$\alpha$&$M_{min}/M_{\sun}$&$M_{max}/M_{\sun}$&Mass Formula&$N_{\star} /V \rho$\\
\hline	
Thin Disk  	&1.6& 0.01 & 1 &$(3.98 - 2.98f)^{-1.66}$ & 3.31\\
Thick Disk   	&0.5& 0.01 & 10 &$(0.316 + 2.846f)^2$ & 0.27\\
Stellar Halo   	&0.5& 0.01 & 10 &$(0.316 + 2.846f)^2$ & 0.27\\
Bulge   	&2.35& 0.7 & 10 &$(22.387 - 22.343f)^{-0.741}$ & 0.593\\
\hline
\end{tabular}
\end{table*}

\subsection{Random Stellar Encounters Using a Galactic Model}
\protect\label{sec:RSE}
In their paper, Robin et al. (2003) gave predicted star distributions in different components of the Galaxy. Their model includes the bulge, thin disk, thick disk and stellar halo
and gives star densities for each depending on Galacto-Centric position (See Tab.2 and 3 of Robin et al. 2003). They also give an initial mass function (IMF) for these different regions (See Tab.1 of Robin et al. 2003).

To find the number of stars in a specific volume we use the following analysis. We are given an IMF $dn(M)/dM \propto M^{-\alpha}$ where $\alpha$ is given in Tab.\ref{tab:reg} for the different components of the Galaxy. To get the number of stars per unit volume between a lower and upper set of mass limits we simply integrate the IMF with respect to
mass such that 

\begin{equation}
\label{eqn:numden}
{N_{\star} \over V} = n = {A[M_{max}^{1-\alpha}-M_{min}^{1-\alpha}]\over 1-\alpha}
\end{equation}
and similarly the mass per unit volume (density) can be found such that
\begin{equation}
\label{eqn:A}
{M \over V} = \rho = {A[M_{max}^{2-\alpha}-M_{min}^{2-\alpha}]\over 2-\alpha}
\end{equation}
Since we already have a value for the density from the model, we can use Eq.\ref{eqn:A} solved for A substituted into Eq.\ref{eqn:numden} to give us the number of stars expected in any particular location in the Galaxy. We have
\begin{equation}
\label{eqn:N}
N_{\star} = {(2-\alpha)[M_{max}^{1-\alpha}-M_{min}^{1-\alpha}]\over (1-\alpha)[M_{max}^{2-\alpha}-M_{min}^{2-\alpha}]}V\rho
\end{equation}

We then use a Poisson random number generator (PRNG) to choose an actual number of stars to simulate from $N_{\star}$ for the reason that $N_{\star}$
can often be 0.1 etc, therefore, if we calculated $N_{\star}$ = 0.1 ten times without a PRNG, we would simulate no stellar encounters. However, the PRNG will give a probability that a non zero number of stars can be generated each time.

Once we have the number of stars, we need to assign a mass and position to each of them. The following argument gives the masses; a uniform random number, $f$, is generated and
this $f$ corresponds to the number density of stars up to a mass, M, divided by the number density of stars up to a maximum mass, $M_{max}$. So from Eq.\ref{eqn:numden}, $f$ corresponds to

\begin{equation}
f = {n(M) \over n(M_{max})} = {M^{1-\alpha}-M_{min}^{1-\alpha}\over M_{max}^{1-\alpha}-M_{min}^{1-\alpha}}{\rm , thus,}
\end{equation}
\begin{equation}
\label{eqn:M}
M = \left[(M_{max}^{1-\alpha}-M_{min}^{1-\alpha})f+M_{min}^{1-\alpha}\right]^{{1 \over 1- \alpha}}
\end{equation}
For the different components of the Galaxy, the specific star density and mass function from Eqs.(\ref{eqn:N}) and (\ref{eqn:M}) respectively are given in Tab.\ref{tab:reg}.

Starting with the microhalo's velocity, $V_{mh}$ and the time-step, $\Delta$t, we know the path length, $l$, it travels in one time-step ($l = V_{mh} \times \Delta t$). Therefore, we can approximate the
volume of the region of interest by choosing some area within which stellar encounters are likely to be significant (for instance a circle of radius 6pc would seem more than
adequate from Fig.\ref{fig:safecrit}.

The star velocities are calculated according to ${\bf V_{\star}}=\frac{V_{rot}}{R_c}(-y_c,x_c,0)$, where $V_{rot}$ = 220km/s and $x_c$, $y_c$ and $R_c$ are the cartesian distances between the COM of the microhalo and the centre of the Galaxy. It is less than ideal since it is considering mainly disk stars, but the relative velocity of the star and halo is not of major importance and
as we shall find, it is almost entirely disk crossing that severely strips the microhalos. There is a minor complication in that Eq.\ref{eqn:imp} requires the relative velocity between the star and microhalo to be a symmetry axis. This transformation is discussed in the appendix and using the results we can transform any position in the unprimed frame into the primed frame by operating on the position vector with a transformation matrix. We also operate on the velocities of
the microhalo particles by this matrix to get velocities in the primed frame. From here we can set up the grid of random stars in the {\bf $y$'}-{\bf $z$'} plane each time step and use a trivially modified
Eq.\ref{eqn:imp} to add the impulses from the star grid. Now we simply take the primed positions and velocities and operate on them with the inverse of the matrix W
(Press et al. 1986), re-add on the core positions and we have simulated random impacts. Notice {\bf $x$'} changes with each time step as the velocity changes with time.

\subsection{Choices of COM}
\protect\label{sec:COCOM}
A complication comes from using the COM as the core of the microhalo, where the core is the centre of the bound particles. The COM is the same as the core for a bound microhalo, however as particles leave and are spread further out in streams, the COM becomes a poor statistic for the core of the microhalo (which we are interested in). When we use an incorrect position of the core in our simulations, we begin to see features in the estimate of bound fraction; such as dipping at pericentre when the velocities are spread and a rapid and premature dive toward being completely destroyed as fully bound particles' velocities relative to the mean velocity of all particles is larger than the local escape velocity. This is clearly shown in Fig.\ref{fig:conver} where we see the blue line dip sharply at pericentre and predict all particles to be unbound before the corrected COM.

Another obvious choice is the median, but this is co-ordinate system dependant. We used a weighted measure of the centre of mass (we call the core) weighted by the square of the particle's potential for all six coordinates of phase space. For phase space $\mu$, any component of the phase space vector ($x,y,z,v_x,v_y,v_z$) we use
\beq
\mu_{core}={\sum_{i=1}^N \mu_i \phi_i^2 m_i \over {\sum_{i=1}^N \phi_i^2 m_i} },\qquad \mu=x,y,z,v_x,v_y,v_z
\eeq
Where $\phi$ is the potential experienced by a particle due to the gravitational influence of all the other particles, without the contribution of the Galaxy. The advantage is that we penalise the escaped particles at large distances where $\phi_i^2 \approx ({Gm \over r_i})^2 \propto {1 \over r_i^2}$. So compared to the COM, $\mu_{core}$ is less sensitive to the extent of the microstream and you can see from Fig.\ref{fig:conver} that it has staggering results.

\section{Results}
\protect\label{sec:res}

\begin{figure*}
\def\subfigtopskip{0pt}    
\def\subfigbottomskip{4pt}
\def\subfigcapskip{1pt}
\centering

\begin{tabular}{cc}
\subfigure[]{\label{fig:orbit28}
\includegraphics[angle=0,width=8.5cm]{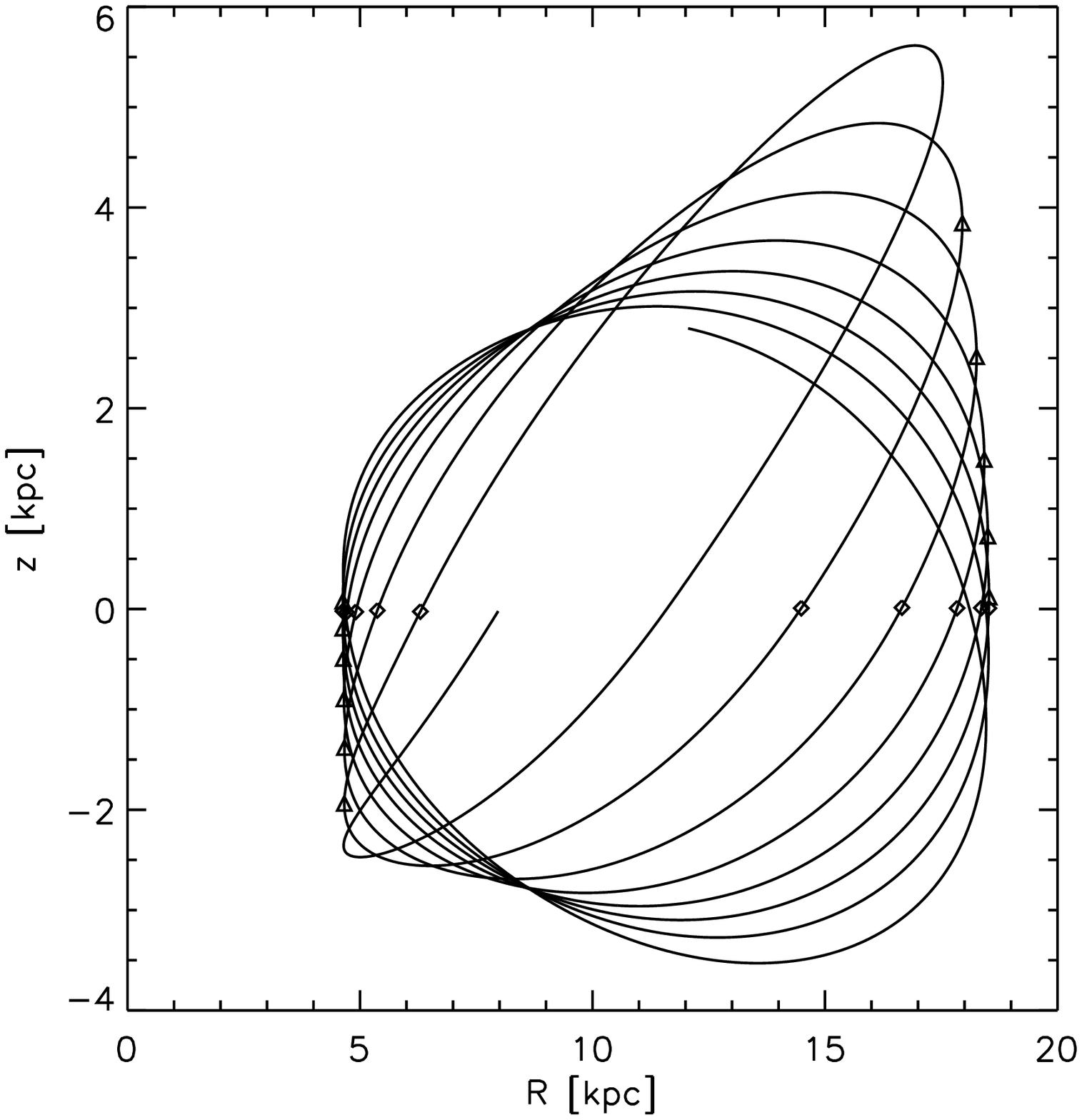}
}
&
\subfigure[]{\label{fig:m3}
\includegraphics[angle=0,width=8.5cm]{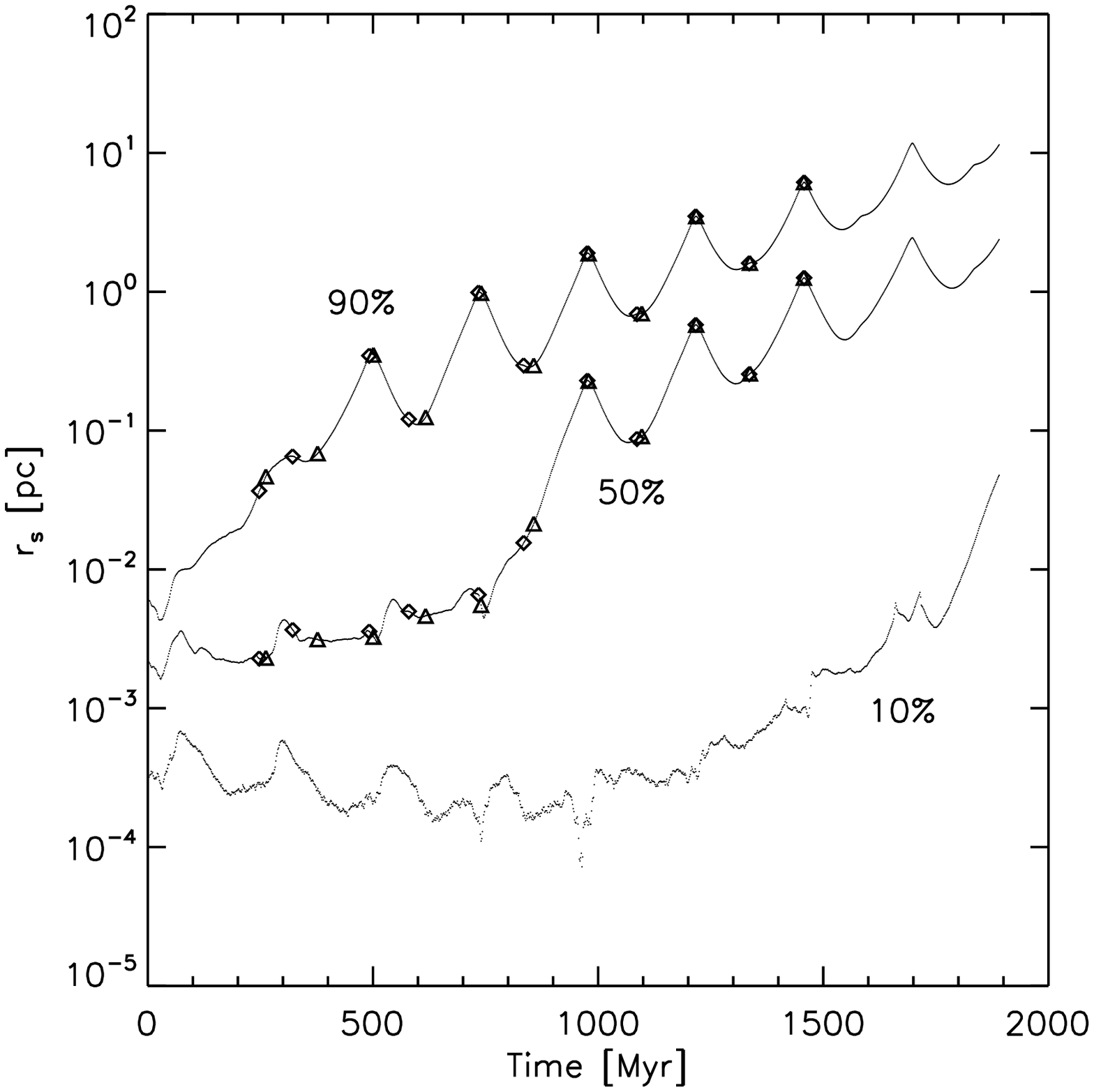}
} \\
\subfigure[]{\label{fig:m2}
\includegraphics[angle=0,width=8.5cm]{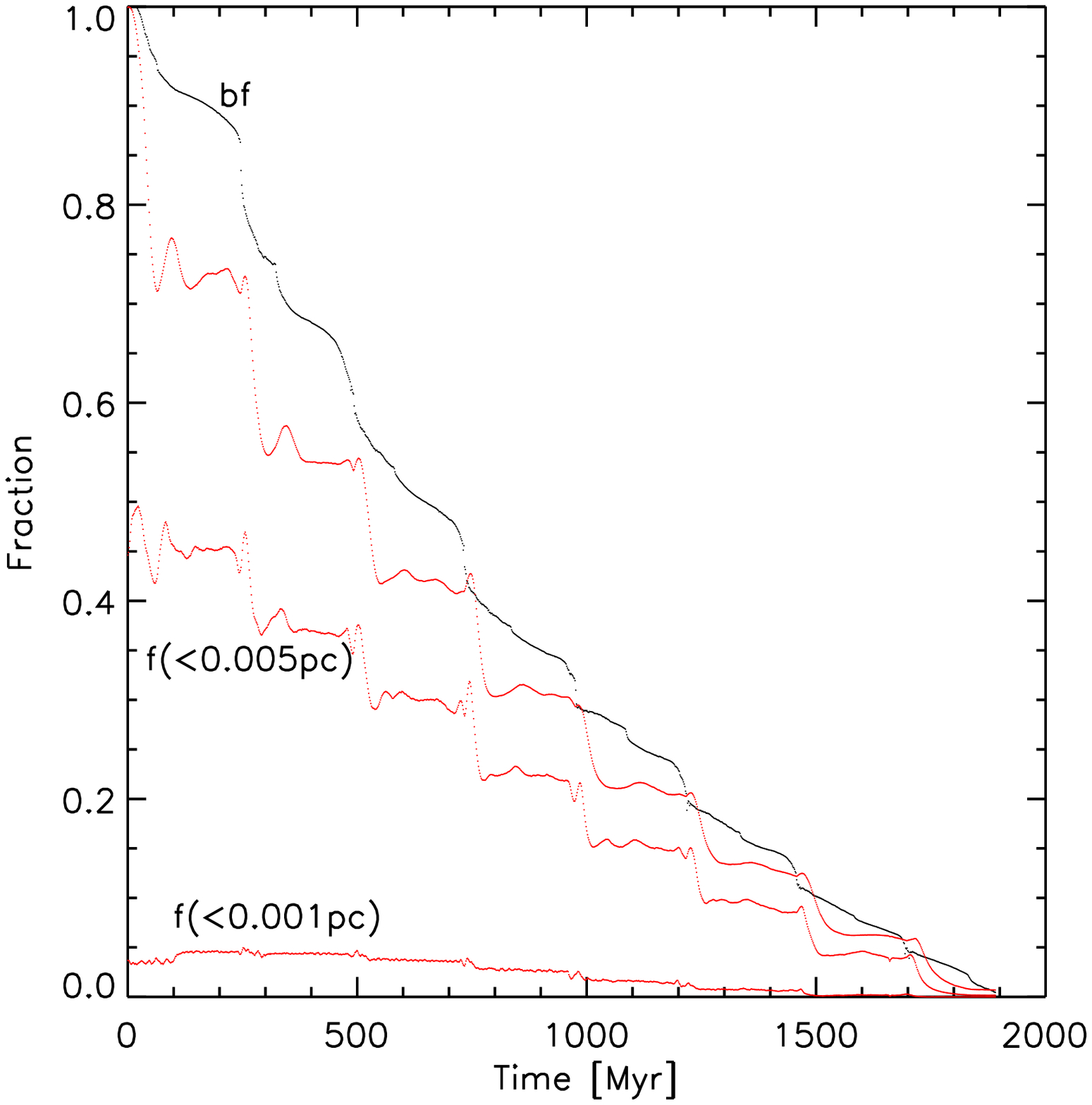}
}
&
\subfigure[]{\label{fig:centesc}
\includegraphics[angle=0,width=8.5cm]{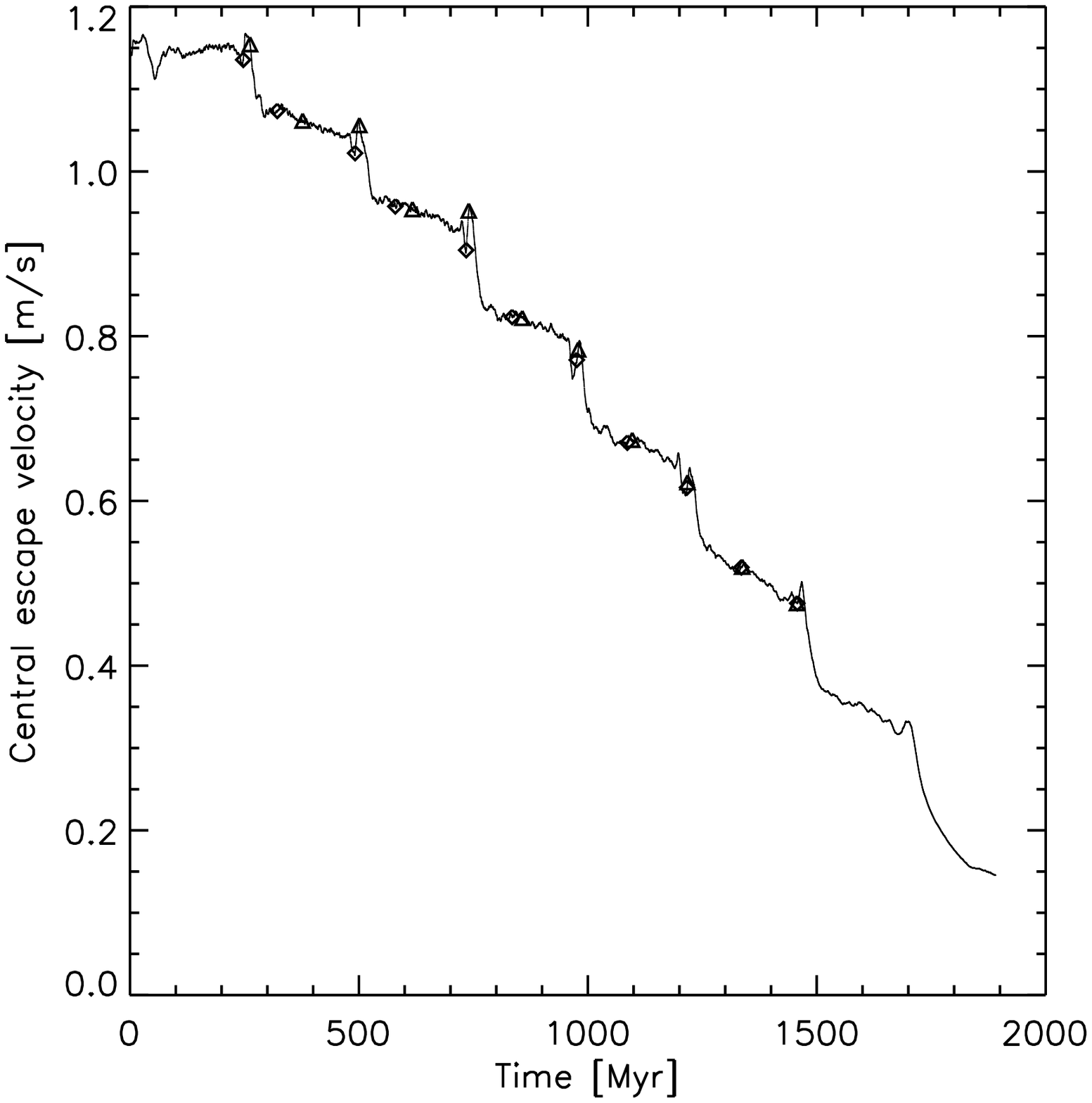}
} 
\end{tabular}
\caption{(a) Shows the orbit of a sample microhalo in the R-z plane during 2Gyr. The disk crossings (diamonds) along with pericentre and apocentre (both with triangles) are marked. (b) Shows the radius of the shells containing 10\% 50\% and 90\% of the microhalo particles. The markers from panel (a) are overplotted which clearly demonstrate that the peaks in the the shell radii correspond to the pericentre of the orbit and the troughs to the apocentres. (c) Shows the bound fraction as a function of time (black line) and also the fraction of particles within three shells of radius $r_s$ = 0.001, 0.005 and 0.01pc (red lines). (d) Shows the central escape velocity plotted against time with the markers from panel (a) overplotted. The minor dip before the staircase drop is during disk crossing and the minor peak is at pericentre.}
\end{figure*}

\begin{figure}
\includegraphics[angle=0,width=85mm]{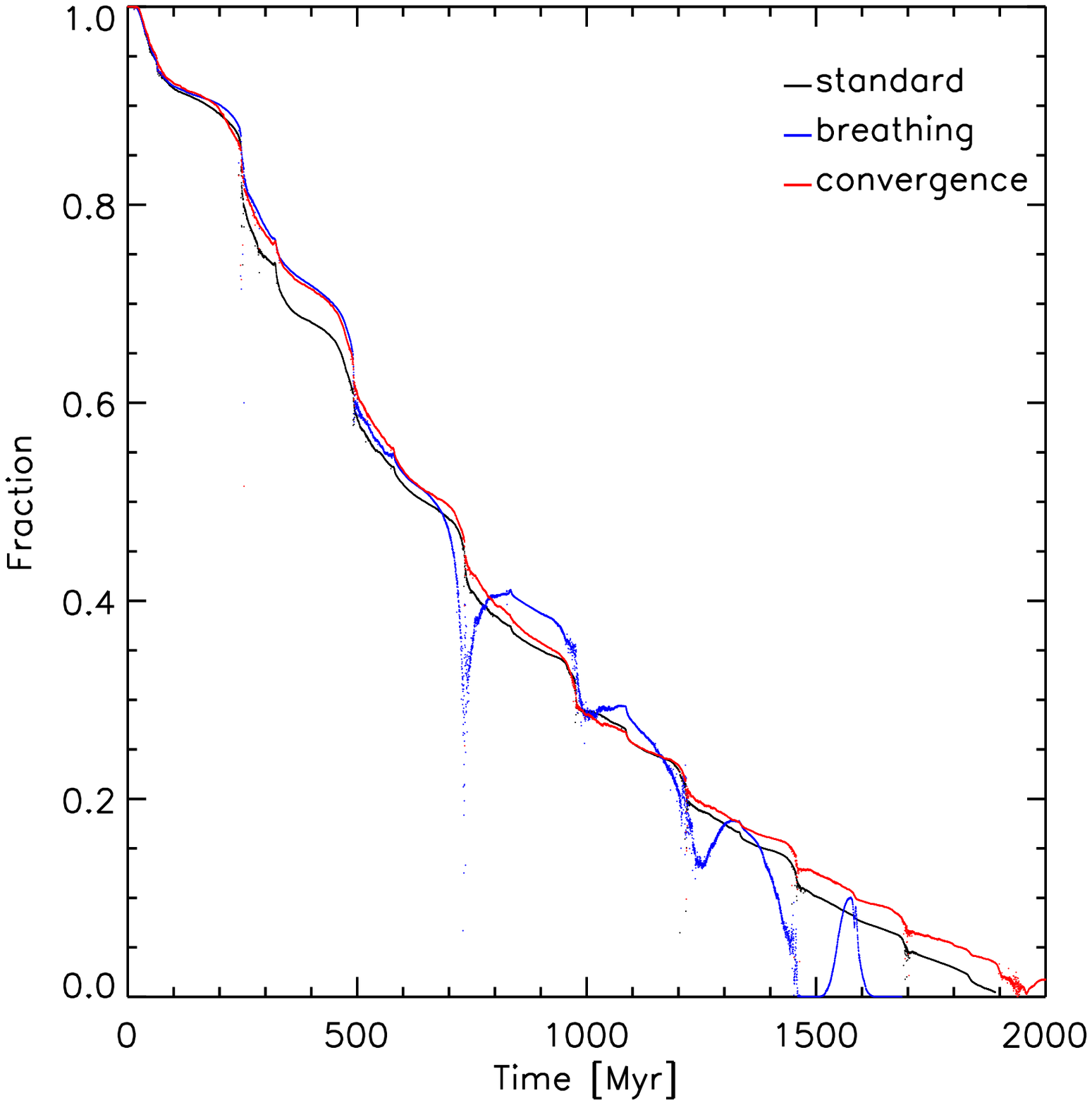}
\caption{The solid black line shows the bound fraction plotted against time exactly as per fig.\ref{fig:m2}. The red line shows the bound fraction from an identical simulation but with only $10^4$ particles instead of $10^5$. The blue line shows an identical simulation to the black line with the same number of particles except that it uses the centre of mass to re-centre the particles instead of the weighted COM as explained in \S\ref{sec:COCOM}.}
\protect\label{fig:conver}
\end{figure}

We used these simulations to investigate the survival probabilities of microhalos depending on the paths they have followed in the Galaxy. In particular, we wanted to know which orbital paths (phase space) are unlikely to still have unspoiled microhalos.

To do this we ran 29 simulations for the TA and RA microhalos with energy and angular momentum which can be seen in Fig.\ref{fig:EJ}, using the random stellar encounters code. We began all simulations at R=8kpc, z=0 and velocities V and $V_z$ as can be seen in Fig. \ref{fig:vzv}, and ran them until the microhalos had less than 5\% of their original bound
particles left. We don't know the detailed distribution of microhalo orbits through the Galaxy, so these orbits are simply a random distribution and the survival times are only a function of orbital properties. Diemand et al. (2005) did not find self-consistent orbits and their comments on the current distribution in the Milky Way are major extrapolations since microhalo orbits would be seriously affected by galaxy formation.

There are certain complications when estimating an instantaneous bound fraction and we used several measures to ensure the microhalo was destroyed. The bound fraction as a function of time is calculated similarly to \S\ref{sec:coswia} by using the potential experienced by each particle purely from the other microhalo particles i.e. offsetting the galactic potential of Eq.\ref{eqn:galpot1}. Then we can say the instantaneous bound fraction consists of all particles with velocity relative to the velocity core less than the local escape velocity defined by Eq.\ref{eqn:esc}.

We used several other estimates of bound fraction to confirm the estimates. The obvious test of bound fraction is whether the particles are actually leaving the microhalo. If they are and we see no particles left inside certain radial shells, we can be sure the microhalo is disrupted. For our simulations we used three shells 10\%, 50\% and 100\% of the original $r_v$. We also considered the radii of the shells, $r_s$ containing the inner 10\%, 50\% and 90\% of the particles. When these radii begin to diverge it means this shell of particles has escaped. The final measure is the central escape velocity. The central escape velocity will become arbitrarily small as the bound fraction decreases and the way it falls should mimic the number of particles within the shells.

\subsection{Analysis of sample orbit}
\protect\label{sec:samporbit}

The important aspects of each orbit are how often the microhalo crosses the disk and the radius from the Galactic centre these crossings occur at (i.e. pericentre and apocentre). The microhalo will survive for longer if it spends long periods of time outside the disk and crosses at large radii. Another aspect is whether the microhalo is co-rotating with the disk or counter-rotating (prograde or retrograde). From Eq. \ref{eqn:imp} we see that the impulse is proportional to the inverse of the relative velocity, therefore, a microhalo on a prograde, planar orbit will have a lower relative velocity than a retrograde orbit. As a result it would be destroyed more quickly.

For a sample orbit (Fig.\ref{fig:orbit28}), we plot the bound fraction against time along with several additional indicators of bound fraction. Notice that the microhalo has a pericentre of 5kpc and apocentre of 18kpc (the markers show disk crossings, pericentres and apocentres). This means that the microhalo will not be affected by its apocentre disk crossings as they are beyond the extend of the thin disk where the bulk of the stars are located. However, the pericentre crossings should strip mass during each encounter. This is exactly what we see in Fig.\ref{fig:m2}. The bound fraction and the fraction of particles within the initial virial radius drop like a stair case and each drop is after a disk crossing at pericentre.

The fraction of particles within the shells of radius 50\% of the virial radius also drops in a consistent manner, but the fraction within 10\% remains fairly constant until most of the particles have been stripped (this can also be seen for another sample orbit in Fig.\ref{fig:streams}b) and all 4 fractions converge when the entire microhalo is destroyed. Whether the inner 5\% of particles is ever destroyed in a real microhalo is an unanswered question and requires higher resolution simulations with variable mass and time steps to be confirmed.

In Fig.\ref{fig:m3} we see three shells which enclose different numbers of particles. The shell enclosing 90\% of the particles decouples almost immediately and increases monotonically except that it oscillates through pericentre and apocentre. The shell enclosing 50\% of the particles detaches at around 500Myr and this is when the bound fraction drops from $\sim$50\% to 40\%. The shell enclosing 10\% of the particles is last to begin to increase. In fact, Fig.\ref{fig:m3} clearly shows that the shells are reasonably constant in radius for the 50\% and 10\% shell before the particles outside this shell escape. Notice that the markers for pericentre and apocentre lie at the crests and troughs of the shell radii respectively as they fluctuate. It is at pericentre that the shells have maximum extent as we might expect as it is at pericentre that the microhalo moves fastest and can thus be spread.

In Fig.\ref{fig:centesc} we plot the central escape velocity as a function of time. This clearly drops in response to the disk crossings as particles are escaping. The markers highlight the feature just before the staircase drop, where the escape velocity takes a small dip (when it crosses the disk) and then rises to a small maximum (at pericentre).

\subsection{Dark Particle Diffusion}
\protect\label{sec:diff}
When crossing a complex star field (i.e. the disk), the microhalo can encounter thousands of stars within the volume given by \S\ref{sec:RSE} in every time step. Most of these stars, if we project them all on to a circle, are too distant to have a measurable effect, however, the stars that pass near the microhalo compete with one another in the single time step as to which direction they pull the microhalo particles. Thus there is always a net velocity shift each time-step and this puts each particles' velocity vector on a random walk. What happens then is that some particles can temporarily receive escape velocity only to get a kick that drops its velocity back below escape.

Even in disk crossings at (R$>$3kpc) critical impacts (see Fig.\ref{fig:safecrit}) are infrequent and the particles' velocities tend to perform small random walks around their mean position until they receive a large kick and then perform a small random walk around the new velocity vector. They can still recover a velocity less than escape, but the larger the kick, the less likely they are to recover orbital velocities. This is especially true if the encounter is penetrating because the particles are not all kicked in the same direction and the symmetry cannot be recovered. 

\begin{figure}
\includegraphics[angle=0,width=85mm]{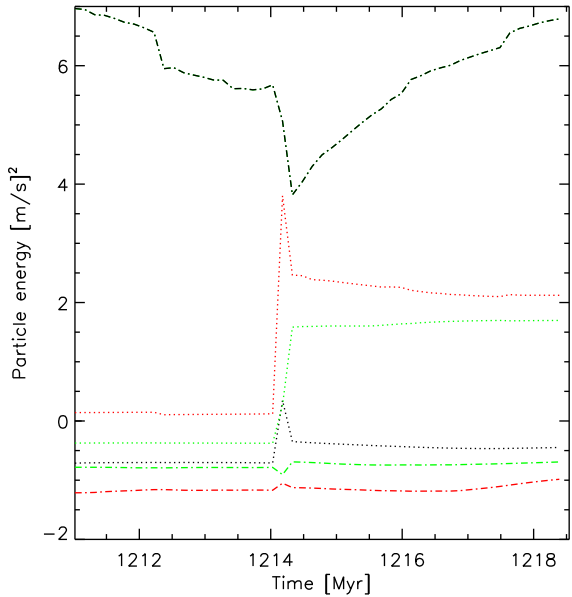}
\caption{Shows the drift in energy of six particles as the second sample microhalo orbit (cf. Fig.\ref{fig:streams}a crosses the disk between 1211 and 1219 Myr. An escaped particle has energy $E>0$. The strong kick received from a close encounter is evident at 1214Myr and the time-step following. The six lines appear flat, but they have small perturbations each time-step that compound. Notice the two most strongly bound particles receive only a minor kick as these particles are likely to be very close to the core and are thus the most difficult to remove.}
\protect\label{fig:diff}
\end{figure}

\begin{figure*}
\def\subfigtopskip{0pt}    
\def\subfigbottomskip{4pt}
\def\subfigcapskip{1pt}
\centering

\begin{tabular}{ccc}
\subfigure[]{
\includegraphics[angle=0,width=5.0cm]{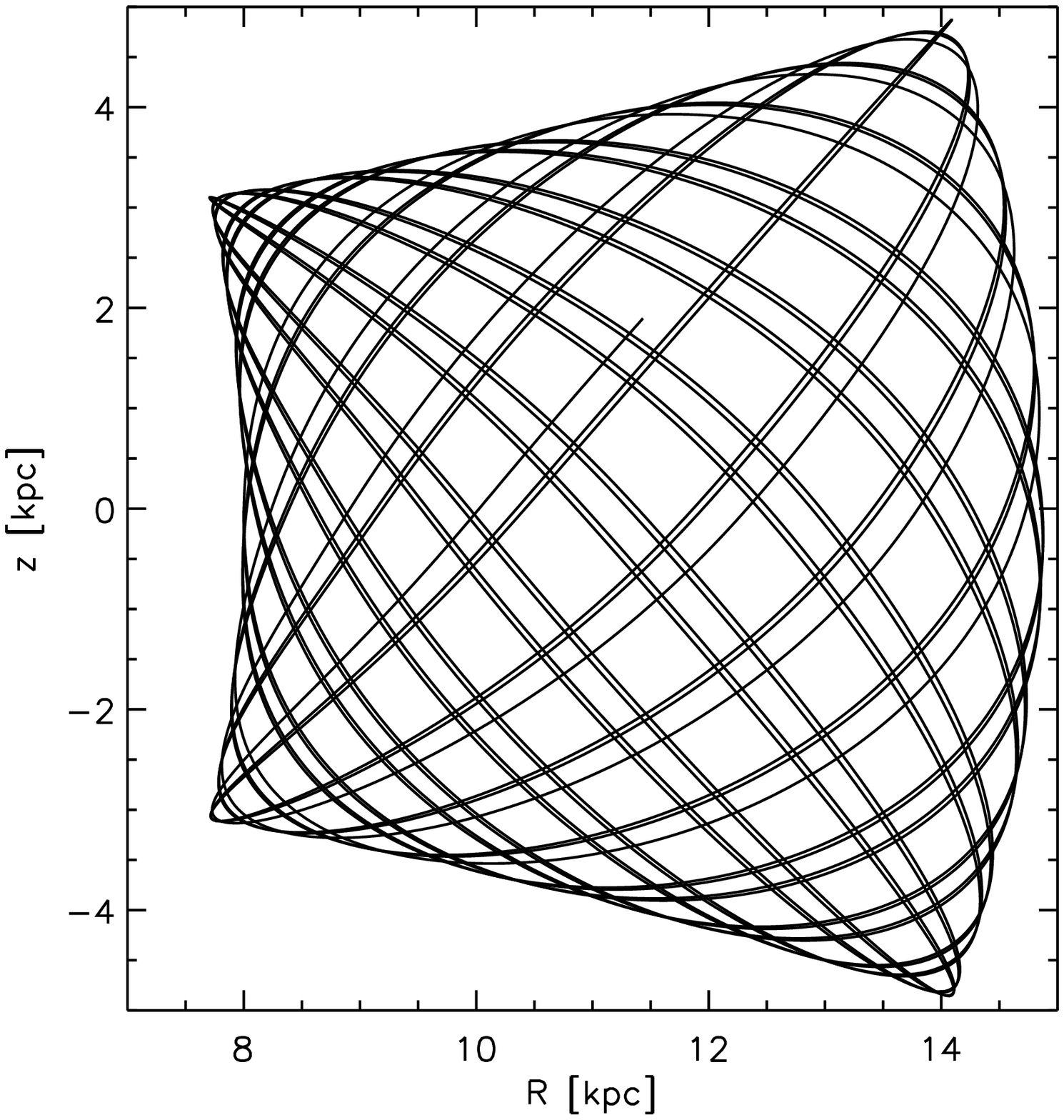}
}
&
\subfigure[]{
\includegraphics[angle=0,width=5.0cm]{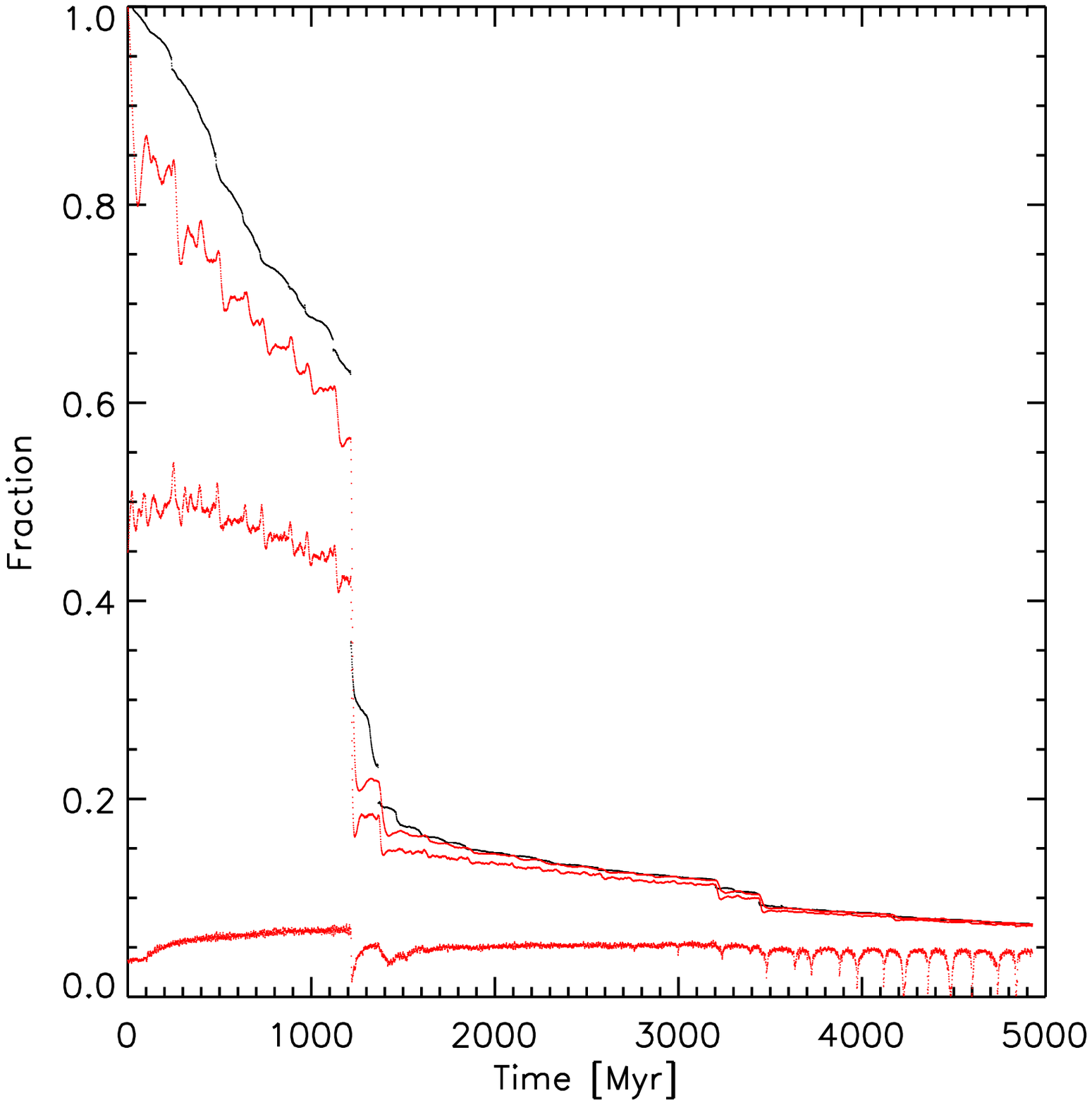}
}
&
\subfigure[]{
\includegraphics[angle=0,width=5.0cm]{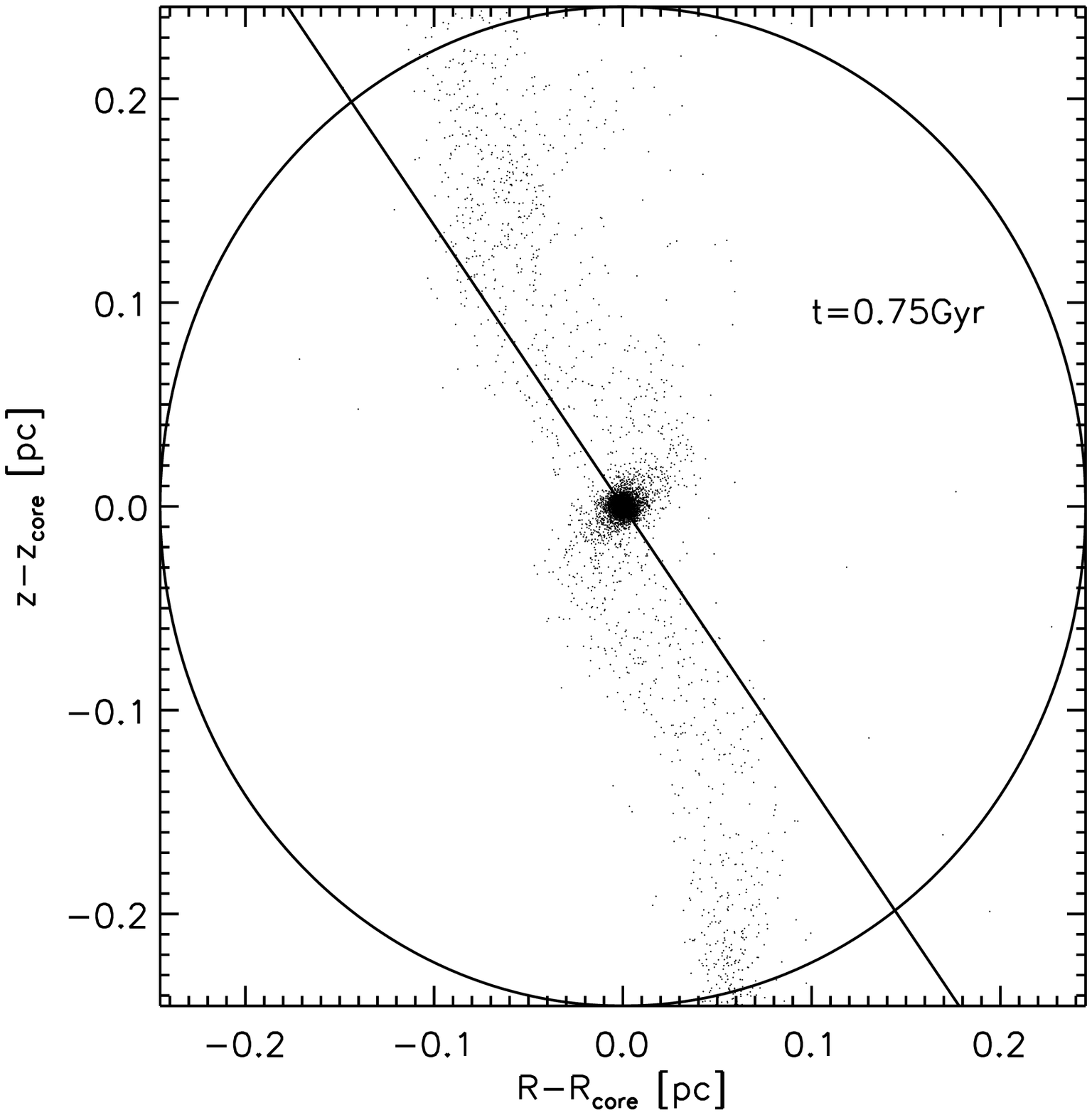}
}\\
\subfigure[]{
\includegraphics[angle=0,width=5.0cm]{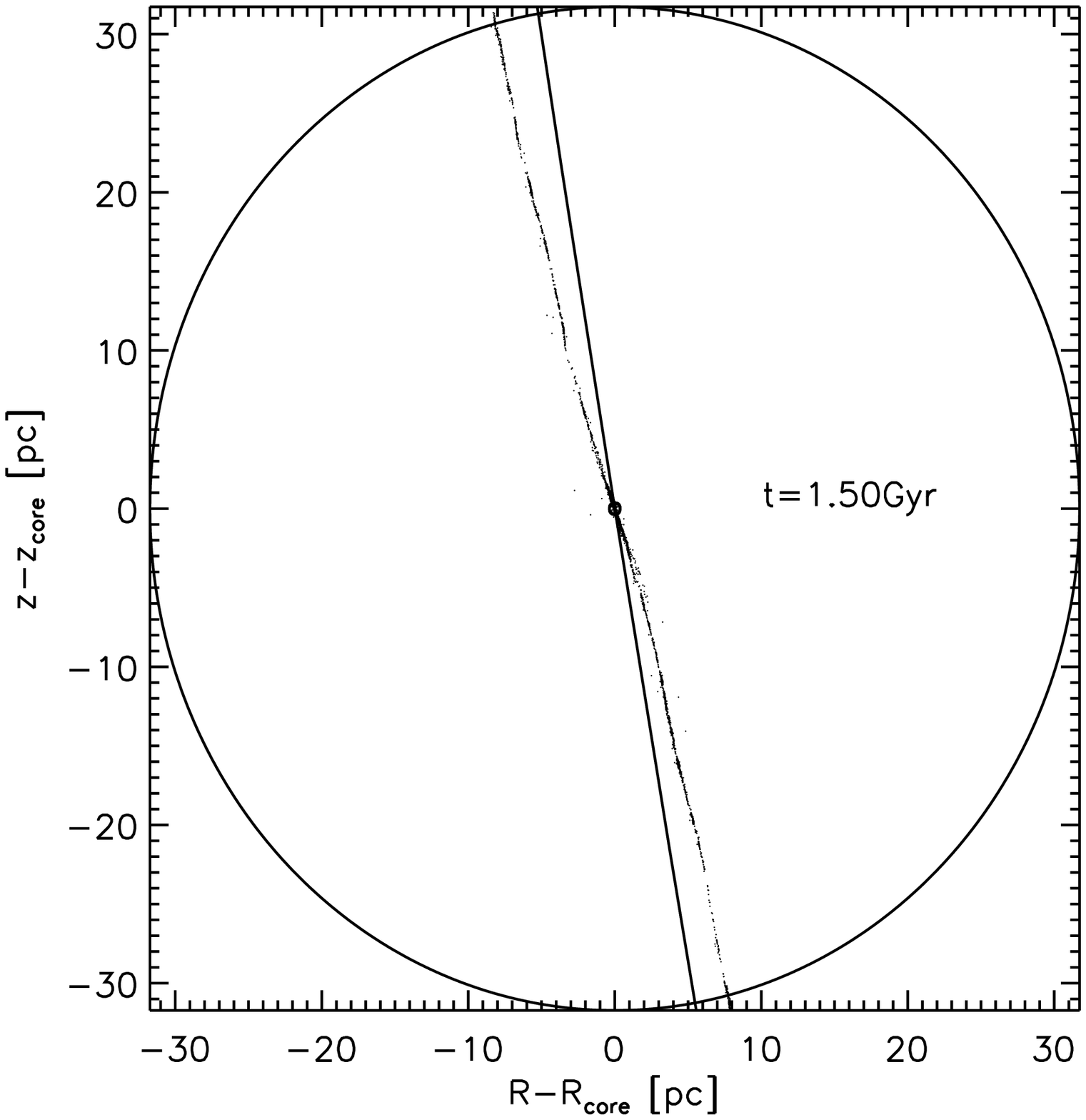}
}
&
\subfigure[]{
\includegraphics[angle=0,width=5.0cm]{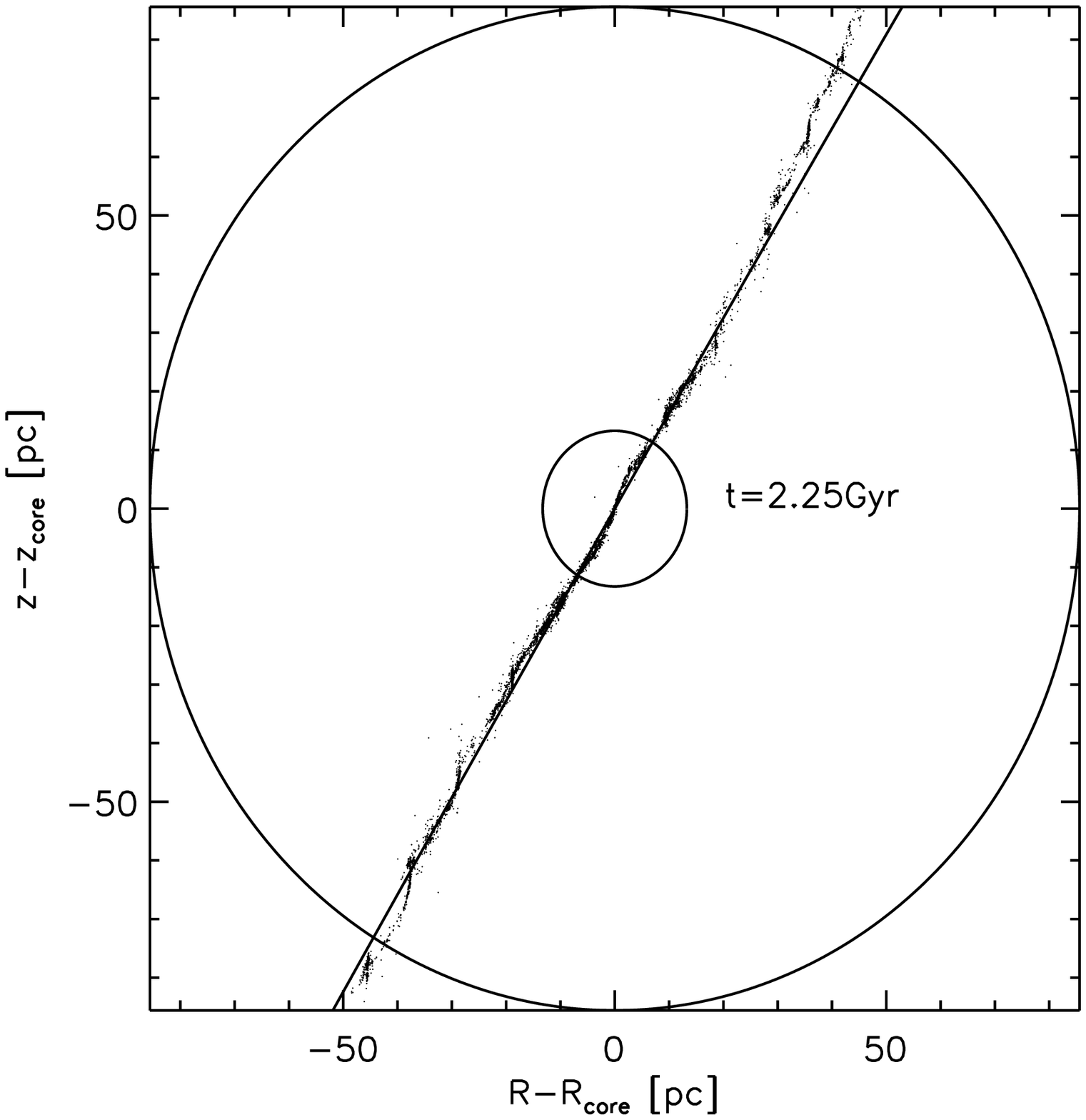}
}
&
\subfigure[]{
\includegraphics[angle=0,width=5.0cm]{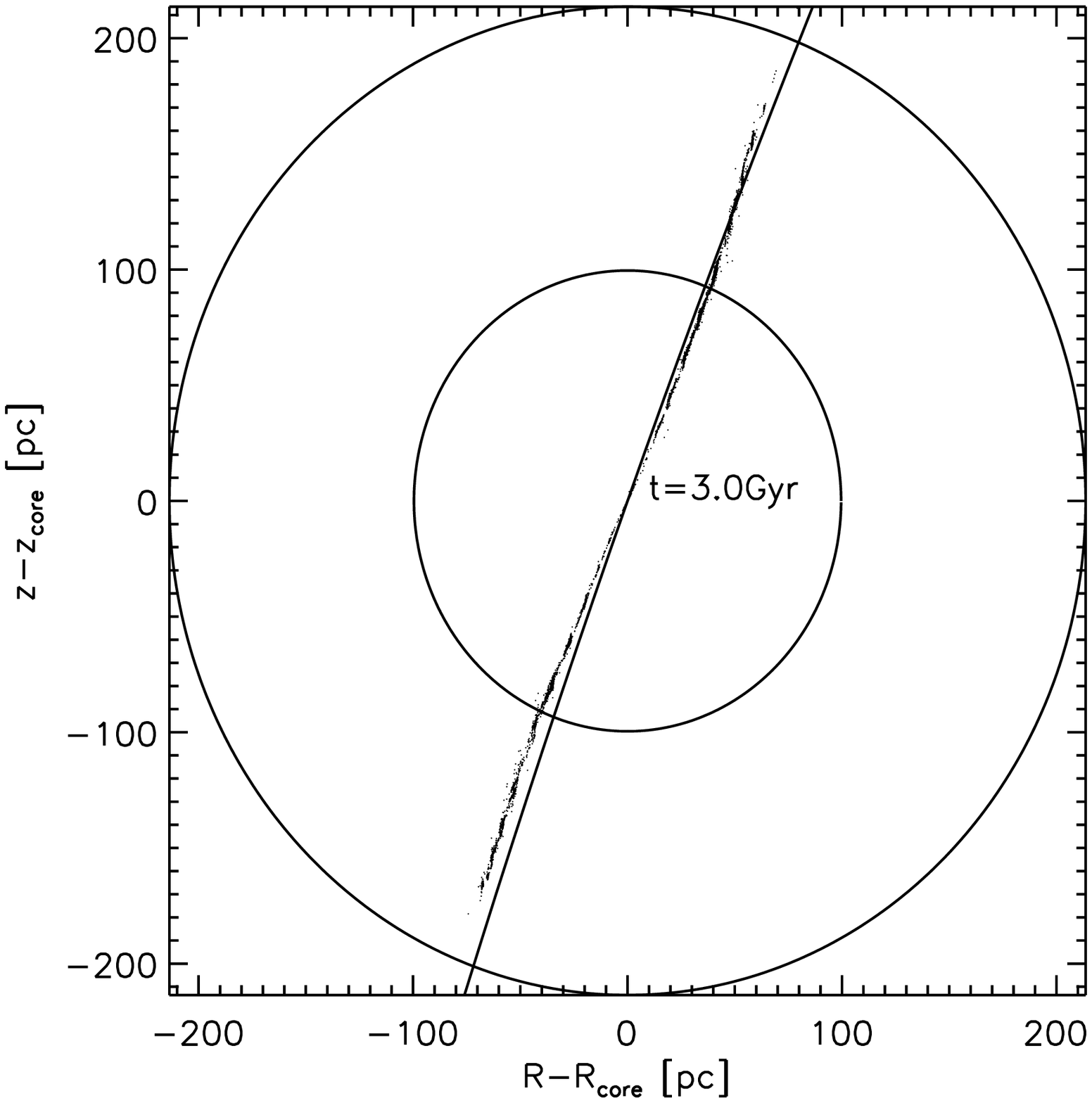}
}\\
\subfigure[]{
\includegraphics[angle=0,width=5.0cm]{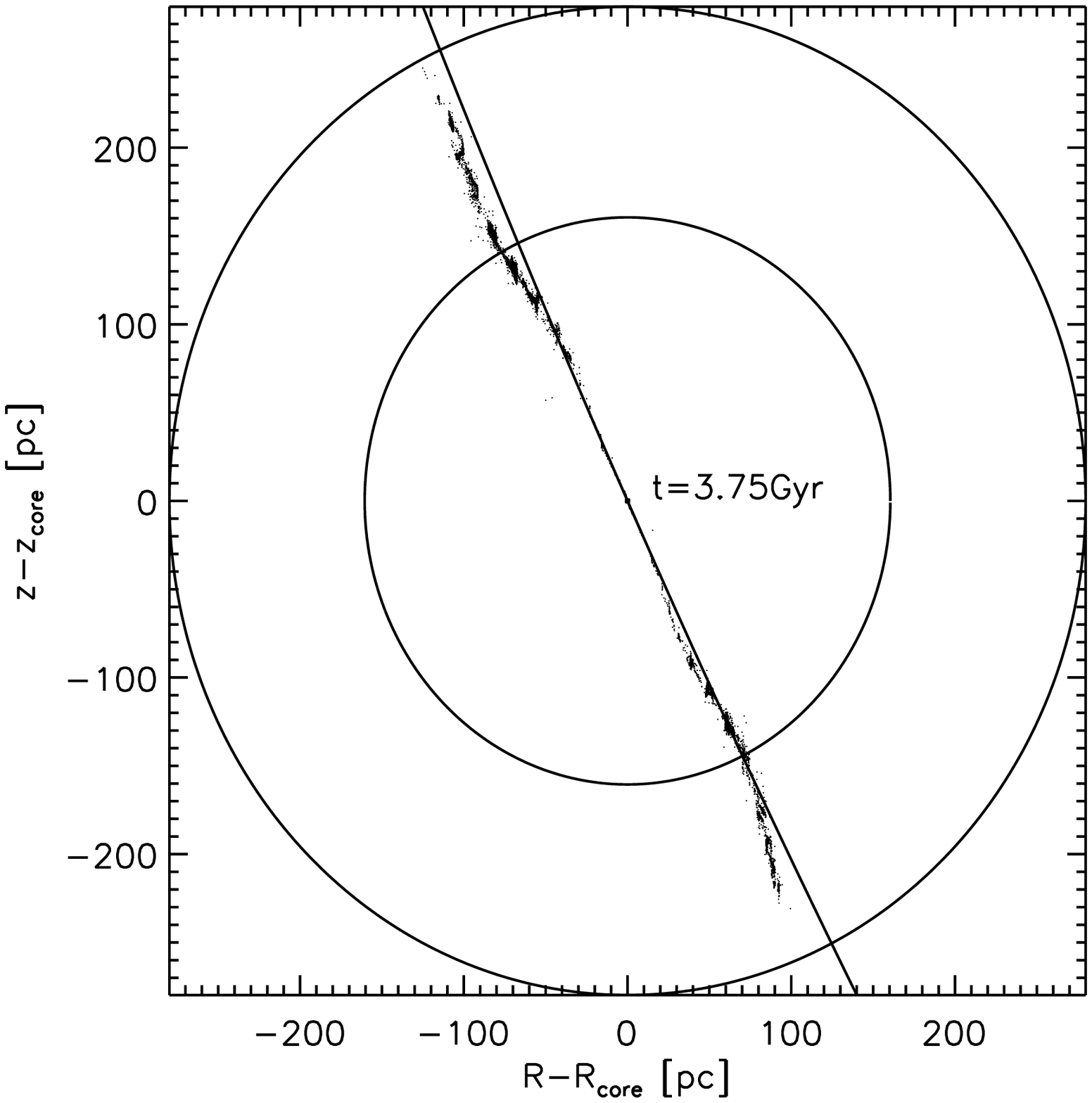}
}
&
\subfigure[]{
\includegraphics[angle=0,width=5.0cm]{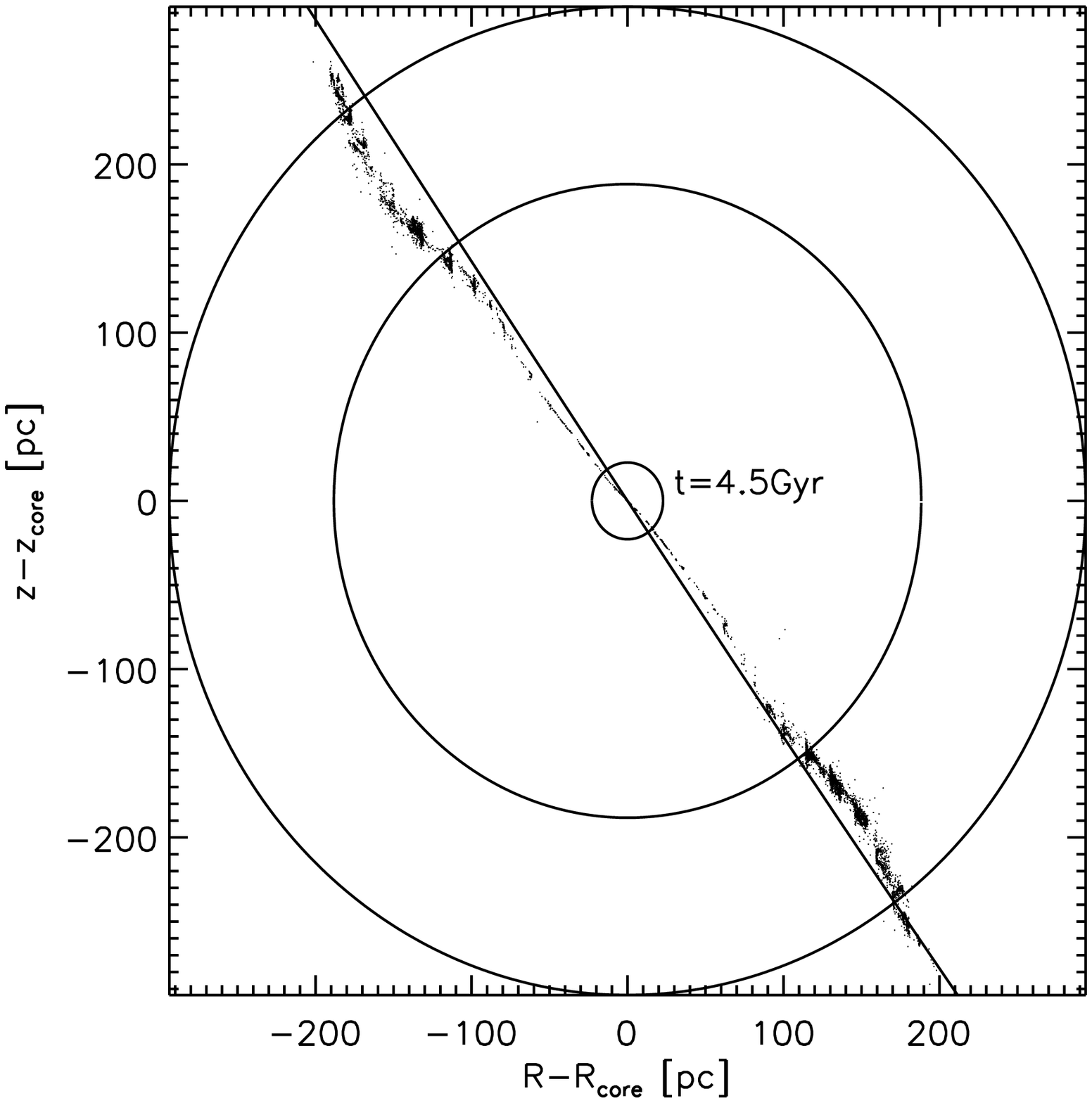}
}
&
\subfigure[]{
\includegraphics[angle=0,width=5.0cm]{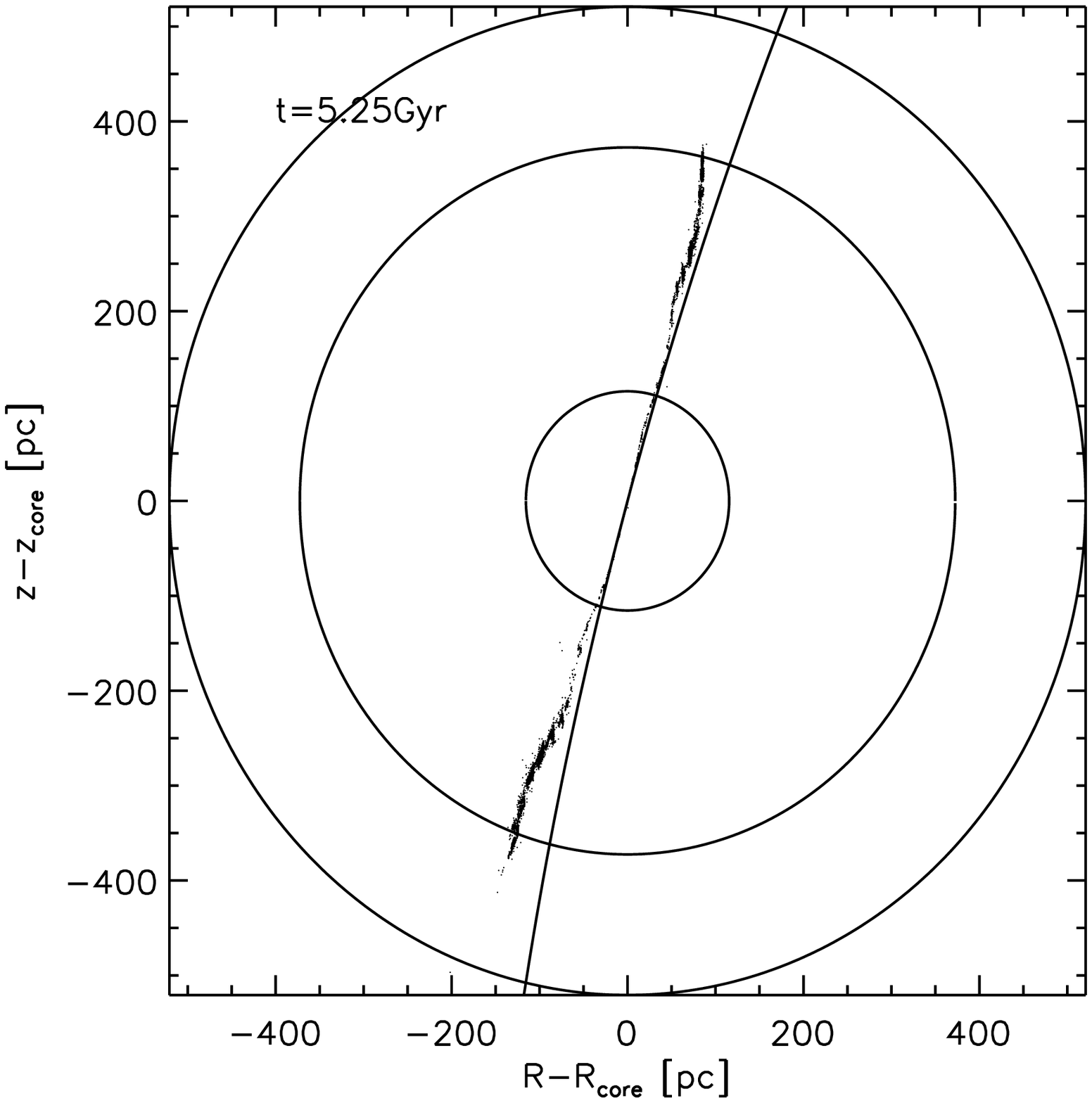}
}
\end{tabular}
\caption{For a new sample orbit shown in (a), we show in (b) the bound fraction (black line) and fraction of particles within three shells (red lines; $r_s$=0.001, 0.005, 0.01pc exactly as in Fig.\ref{fig:m2}). Then in (c)-(i) we show the microstream created after the disruption of the microhalo at different pericentric disk crossings from 750Myr to 5.25Gyr: these crossings all happen near $R=8$kpc.  The shell radii in parsecs enclosing 10\%, 50\% and 90\% of the microhalo particles are shown. This microhalo still retains 5\% of its particles at 6Gyr. The overplotted line is the orbital path of the core of the microhalo. From (a) note that at pericentre the microhalo always crosses the disk from above the plane. In some streams the 10 and 50\% shells are not visible as they are shrouded inside the core. We only show 10\% of the $10^5$ particles to avoid large files. Beyond 4Gyr the number of particles inside the shell $r_s=0.001pc$ shown in (b) is unstable at pericentre.}
\protect\label{fig:streams}
\end{figure*}

It is by this process of diffusion that the particles gradually gain escape velocity during disk crossings aided by the Galactic tides. Once it has left the disk, the lack of perturbations allows the particles which achieved escape velocity to leave and ones that hadn't quite achieved escape can still escape if enough other particles do, allowing the potential well to shallow. This effect can be seen in Fig.\ref{fig:diff} where we see the velocities of six random particles just before and just after the disk crossing from the orbit of Fig. \ref{fig:streams}a. Notice from Fig. \ref{fig:streams}b that the microhalo receives a strong kick during this disk crossing which drops the bound fraction from 65\% to 25\% in one time-step and additionally that there is a second kick in the following time-step which restores around 5\% of the particles. When the particle has energy $E = {1 \over 2} v^2 +\phi>$ 0 it has escape velocity; where $v$ is the particles velocity relative to the microhalos net orbital velocity and $\phi$ is the potential at the particles position imposed solely by the other microhalo particles. Notice that some of the particles gain energy and some lose energy. This depends on the position of the particle with respect to the microhalo core and the star. If it is between the microhalo and the star it will gain energy and the opposite is true if it is on the opposite side.

\subsection{Survival times}

\begin{figure*}
\def\subfigtopskip{0pt}    
\def\subfigbottomskip{4pt}
\def\subfigcapskip{1pt}
\centering

\begin{tabular}{cc}
\subfigure[]{\label{fig:surv50}
\includegraphics[angle=0,width=8.5cm]{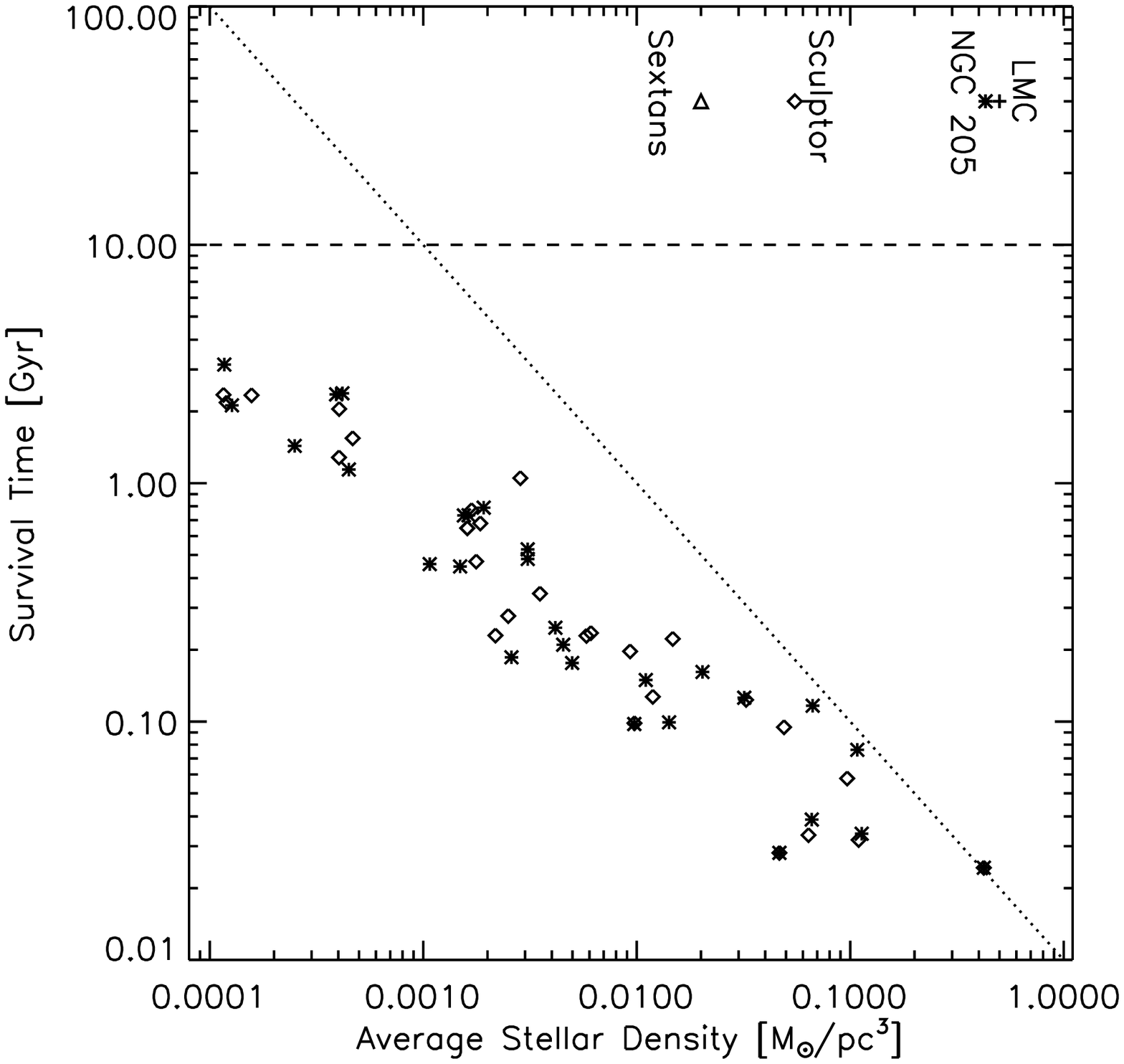}
}
&
\subfigure[]{\label{fig:surv5}
\includegraphics[angle=0,width=8.5cm]{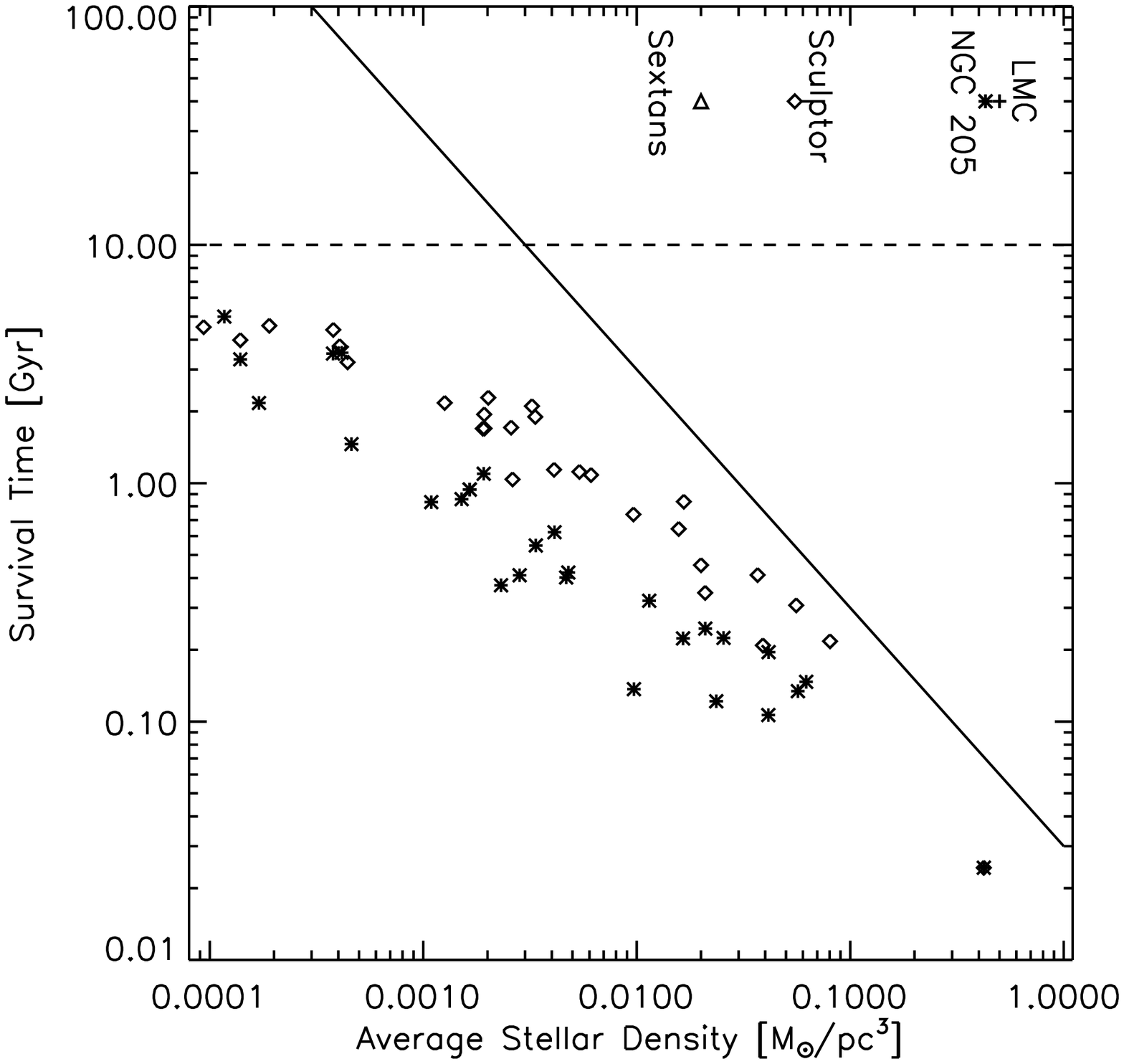}
}
\end{tabular}
\caption{Here the survival time of RA microhalos (asterixes) and TA microhalos (diamonds) for 50\% (a) and 5\% (b) survival is plotted against the average stellar density $\overline{\rho_{\star}}$.  The overplotted line in each plot is the prediction from the semi-analytic model of Zhao et al. (2006). There is a clear trend for TA microhalos to survive at the 5\% level for longer than RA microhalos. This plot suggests there is a critical stellar density below which microhalos should survive a Hubble time i.e. $\sim 10^{-4} M_{\sun} pc^{-3}$. As a point of reference, the {\it central densities} in the Sculptor, Sextans, LMC and NGC 205 galaxies are plotted. It is interesting to note that in the nucleus of M32 the density is $\sim  786M_{\sun} pc^{-3}$ which is far off the scale.}
\protect\label{fig:surv}
\end{figure*}

\begin{figure*}
\def\subfigtopskip{0pt}    
\def\subfigbottomskip{4pt}
\def\subfigcapskip{1pt}
\centering

\begin{tabular}{ccc}
\subfigure[]{\label{fig:vzv}
\includegraphics[angle=0,width=6.0cm]{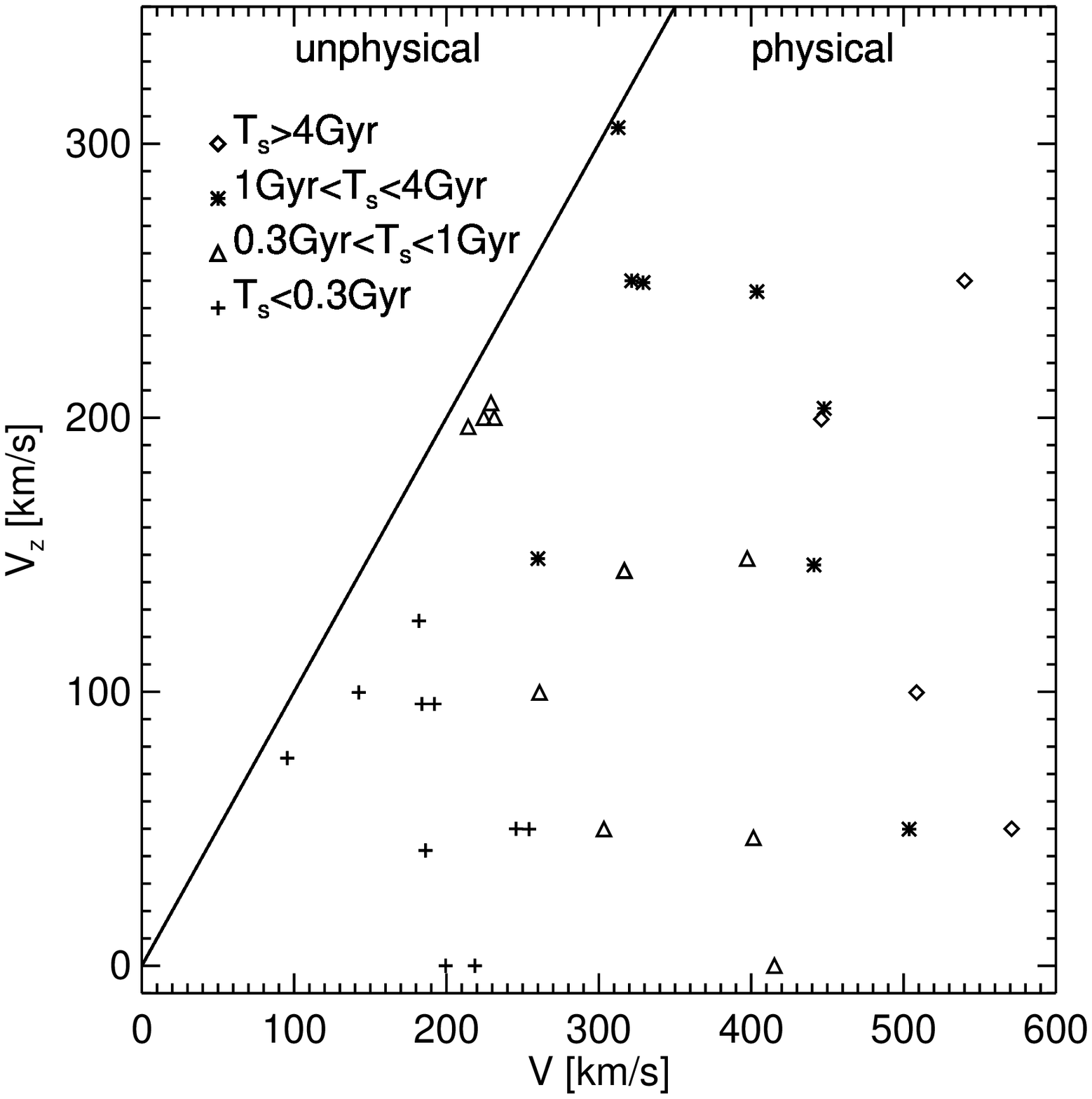}
}
&
\subfigure[]{\label{fig:EJ}
\includegraphics[angle=0,width=6.0cm]{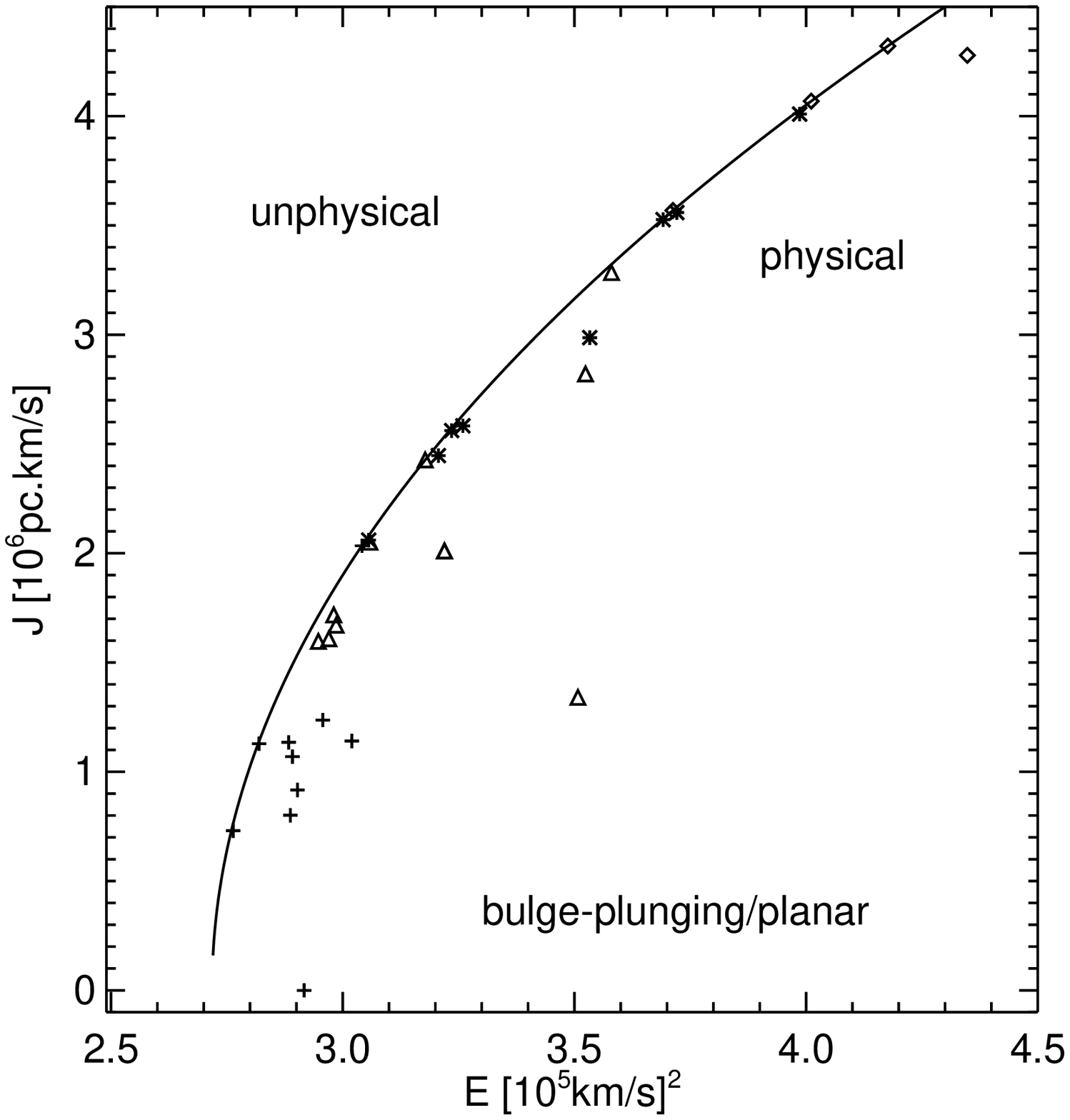}
} 
&
\subfigure[]{\label{fig:EJz}
\includegraphics[angle=0,width=6.0cm]{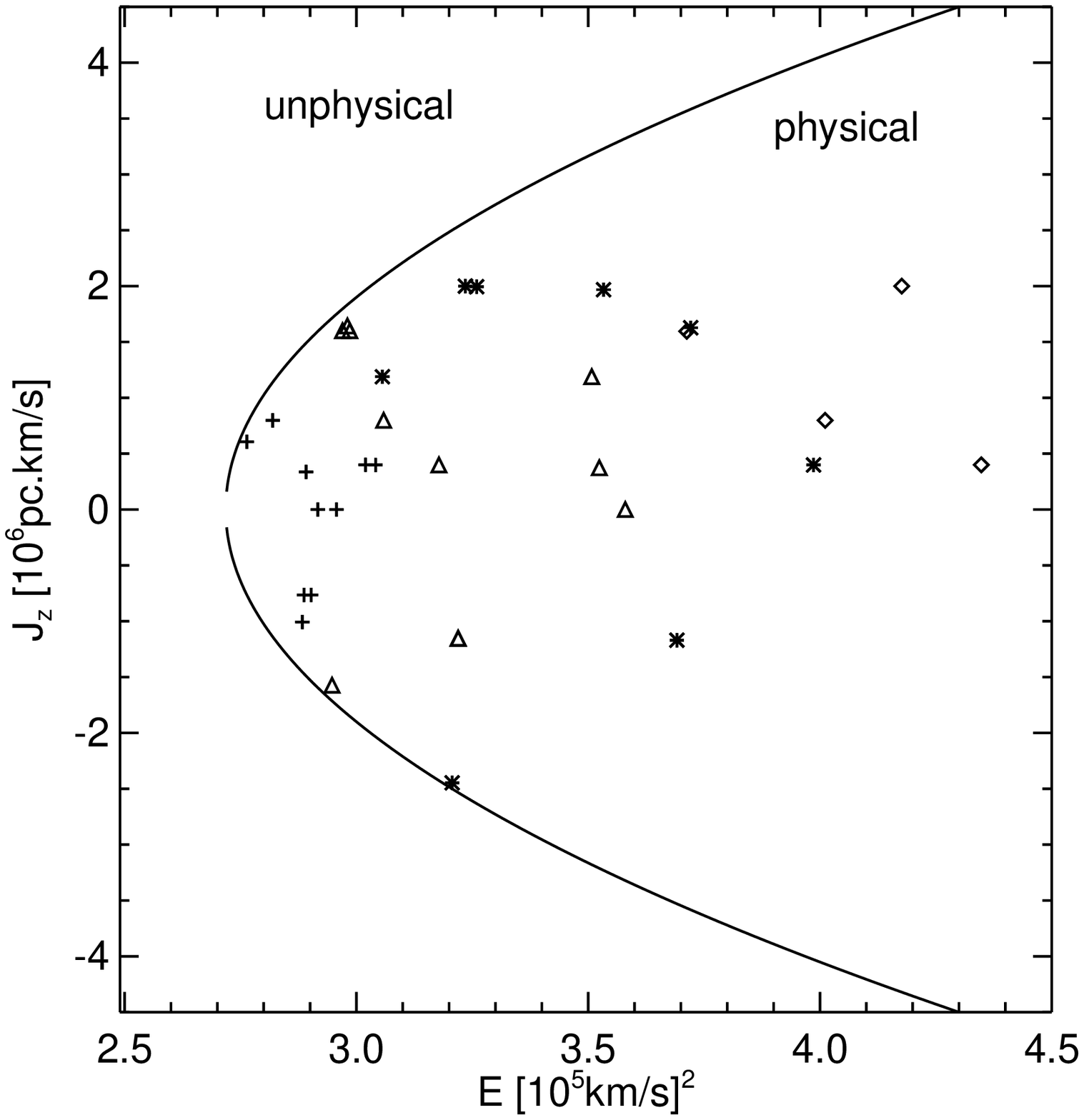}
} 
\end{tabular}
\caption{For all the RA microhalos. (a) Shows the initial $V_z$ vs. V. (b) Shows initial J vs. E and (c) shows initial $J_z$ vs. E, where the different symbols represent different ranges of 5\% survival time. Diamonds are for undamaged microhalos after 4Gyr, crosses, triangles and stars are for $T_S$ $<$ 0.3, 1 and 4 Gyr respectively. Unphysical and physical regions of the plots are highlighted.}
\end{figure*}

In order to make some estimates of the timescales of microhalo survival we have used the time taken for a microhalo to lose 95\% of its particles (i.e. to drop to a bound fraction of 5\%) and to lose 50\%. Additionally, it was found that the local average stellar density encountered by a microhalo on its orbit was a useful quantity as it is a function of pericentre, apocentre, orbital angular momentum, number of disk crossings and where the Galactic tides are strongest. It is defined as

\beq
\overline{\rho_*}={\sum_{i=1}^{n_{ts}} \rho_{*,i} \over N}, \qquad \rho_{*,i}={ \sum_{j=1}^{n_*} m_{*,j} \over \pi a^2 \Delta t V_{mh,i}  }
\eeq

Where $\rho_{*,i}$ is the stellar density encountered at the $i^{th}$ time-step. $n_{ts}$ is the time-step at which the microhalo loses a specified fraction of its mass (95\% or 50\%), $n_*$ is the number of stars simulated in the $i^{th}$ time-step, $m_{*,j}$ is the mass of the $j^{th}$ star, $\Delta t$ is the time-step, $V_{mh,i}$ is the core velocity of the microhalo in the $i^{th}$ time-step and a is the arbitrary radius out to which stars are simulated, chosen to be 6pc.

The 50\% and 5\% survival times are plotted against average stellar density $\overline{\rho_*}$ in Fig. \ref{fig:surv} for TA and RA microhalos. All points seem consistent with a power law relationship between both survival times and $\overline{\rho_*}$, however, it is apparent there is a significant difference between the TA and RA microhalos. They both have more or less the same survival times for 50\% of the particles but the TA microhalos are far harder to bring down to a bound fraction of 5\%. The natural explanation of this is the radial orbits bring particles out to larger radii making the anisotropic microhalos a larger target for the stars.The reason the TA and RA microhalos don't experience the same $\overline{\rho_*}$ in Fig. \ref{fig:surv} is a consequence of the significantly longer survival of TA microhalos allowing them to experience slightly different average densities. The fluctuations in survival time are expected as they can depend on the order in which the microhalos disk cross at pericentre and obviously due to the random nature of the stellar encounters.

Due to the limitations of the tree-code a 1Gyr orbit takes about 50 hours on a standard desktop, so we limited most of our simulations to 4Gyr (comparable to the actual time that the disk stars have been cohabitating with the microhalos). Certain simulations in our run were not destroyed after 4Gyr, so to include them in our conclusions, we interpolated to find the survival time considering the past decline of bound fraction and when the microhalo will cross the disk. This is not possible when the microhalo has a bound fraction of more than $\sim$15\% at 4Gyr as we have not followed the trend of the bound fraction for long enough to project it, so we did not include these orbits in our sample.

\subsection{Phase space survival}

Many authors have noted the higher survival probabilities of microhalos depending on certain orbital parameters. Pericentre, apocentre, orbital energy, magnitude and sign of orbital angular momentum and even internal anisotropy as we have considered. Our simulations all begin at the Galactic position (R,z)=(8kpc,0).

The energy of an orbit is defined in the following way

\beq
E={1 \over 2}\left(v_R^2 + v_{\theta}^2 + v_z^2  \right) + \Phi(R,0) ={1 \over 2}\left(v_R^2 + {J^2 \over R^2}\right) + \Phi(R,0).
\eeq
Where $v_R, v_{\theta}$ and $v_z$ are the orbital velocities of cylindrical co-ordinates where negative $v_R$ is directed towards the centre of the Galaxy, $v_{\theta}$ is tangential and $v_z$ is perpendicular to the disk. $\Phi(R,0)$ is the Galactic potential at radius R in the disk.  The total angular momentum $J$ (approximate integral of motion) and its z-component (conserved) of the orbit is defined as
\beq
J=R\sqrt{v_{\theta}^2+v_z^2},\qquad J_z=Rv_z.
\eeq

For the RA microhalos we have plotted initial orbital $V_z$ vs. V in Fig.\ref{fig:vzv} for the 29 microhalos using four different symbols depending on their individual survival times. This hints at the different regions of phase space which are likely to have surviving microhalos. In Figs.\ref{fig:EJ} and \ref{fig:EJz} we plot the orbital angular momentum J and $J_z$ against energy E for the RA microhalos. The unphysical regions of the plots simply show that for a certain $v$ or energy at a specific distance a microhalo is limited in its maximum $v_z$, J or $J_z$. There is a clear trend for the microhalos on high energy and angular momentum orbits to survive longer. Written another way, microhalos on planar and radial orbits will be rapidly destroyed, whereas increasing tangential velocities will enable them to survive intact for greater timescales.

\subsection{Microstreams in the solar neighbourhood}

If it is the case that survival depends on orbital properties we can begin to think about the microstreams created by the destruction of microhalos. We show in Fig.\ref{fig:streams} the sample microhalo at multiple stages of its evolution. Clearly, the microhalo becomes increasingly filamentary, but even after over 5Gyr , the stream is still a distinct entity that would stand out in the presence of a background thanks to the low escape speeds.

From the discussion of orbital parameters and the microstreams there are clearly many implications for the direct detection signal from dedicated dark matter searches because depending on the velocity of microhalos, especially perpendicular to the Galactic plane and opposing the rotation of the disk, they will have various degrees of concentration. Additionally, it may be the case that most microhalos passing near the Earth have been dispersed into streams and our detector would have a boosted probability of actually passing through a stream or a superposition of many streams. 

\section{Conclusions and Discussion}
\protect\label{sec:conc}
Microhalos of $10^{-6}M_{\sun}$ are likely the first substructures formed in a CDM universe. Those microhalos in dense regions of galaxies would experience strong tidal stripping. Here we presented N-body simulations of microhalos in single encounters with stars as well as in a series of encounters in a realistic Milky Way.

The clear trend is that for orbits close to the solar neighbourhood the survival time falls off as a power law of the average stellar density and the only likely surviving microhalos are ones on orbits with high energy and angular momentum; also the TA microhalos have a slightly better chance of survival than the RA microhalos.
\\
\\
Nevertheless, some cautionary points:
\\
\\
1. We consider only Sun-passing orbits. Those closer to the bulge are likely more severely stripped and more distant ones are less interesting for detection.
\\
\\
2. Our microhalos all have the same total mass of $10^{-6}M_{\sun}$ and fixed initial tidal radius (0.01pc).  In realisty slightly more massive microhalos may endure stellar encounters to greater effect and exist in higher numbers today.  Also the initial tidal radius should vary with the pericenter of the microhalo.
\\
3. The stars in our Galaxy are not formed all at exactly 4 Gyrs ago.  The granularity and the flattening of the galactic potential presumably increased gradually over a Hubble time.

However, our method shows how the impulsive encounters with stars alongside disk shocking and tidal stripping poses a serious threat to the survival of primordial microhalos. The statistical nature of survival that we have demonstrated could be used or at least elaborated upon to predict statistically the phase space distributions of microhalos and morphology in order to make more accurate estimates of annihilation and direct detection signals.

\section*{Appendix}

This appendix shows how to make the co-ordinate transformation required in \S\ref{sec:RSE}. In Eq.\ref{eqn:imp}, we simply had a relative velocity along the $x$-axis. Now we have the case where the relative velocity is a combination of {\bf $x$}, {\bf $y$} and {\bf $z$}. This requires us to transform to
a primed set of coordinates {\bf $x$'},  {\bf $y$'} and {\bf $z$'}, where {\bf $x$'} is the symmetry axis. 
The transformation is relatively straight forward. We find the relative velocity between the microhalo and the star by a simple vector subtraction ({\bf$V_{rel} = V_{mh} -
V_{\star}$)}. Then we assign the direction {\bf x' = $V_{rel}$/}$|V_{rel}|$. No perturbations occur in the plane of {\bf $x$'} due to symmetry. So we need to pick any two mutually orthogonal
directions in this plane to be represented as {\bf $y$'} and {\bf $z$'}. {\bf $y$'} is easy; from the vector dot product, {\bf x'$\cdot$y'} = $(x_1',x_2',x_3')\cdot(y_1',y_2',y_3')$ = 0. So, if we say, for
simplicity's sake, that $y_1'$ = 0 and $y_3'$ is any real number s.t. $y_3'$$\epsilon$(0,1), then $y_2'$ = -$\frac{x_3'}{x_2'}y_3'$. Since $z$' must be mutually orthogonal to {\bf $x$'} and {\bf $y$'}, we simply take the vector cross product of these two vectors to produce {\bf $z$'}. Both {\bf $y$'} and {\bf $z$'} require renormalisation.

If we put these three vectors into a matrix W, s.t
\begin{equation}
\begin{array}{l|rrr|}
    &x_1'&x_2'&x_3'\\
W = &y_1'&y_2'&y_3'\\
    &z_1'&z_2'&z_3'\\
\end{array}
\end{equation}

\section*{Acknowledgements}

\label{lastpage}

\end{document}